\documentclass[twocolumn,showpacs,preprintnumbers,amsmath,amssymb,letterpaper]{revtex4}
\usepackage{graphicx}% Include figure files
\usepackage{dcolumn}% Align table columns on decimal point
\usepackage{bm}% bold math

\usepackage{color}

\newtheorem{theorem}{Theorem}
\newtheorem{lemma}[theorem]{Lemma}
\newtheorem{corollary}[theorem]{Corollary}

\newenvironment{proof}[1][Proof]{\noindent\textbf{#1.} }{\ }

\begin{document}

\preprint{APS/213-QED}

\title{Squeezing Components in Linear Quantum Feedback Networks}% Force line breaks with \\

\author{J.~E.~Gough}
\email{jug@aber.ac.uk}

\affiliation{Institute for Mathematics and Physics, Aberystwyth University, SY23 3BZ, Wales, United Kingdom.}

\author{M.~R.~James}%
\email{Matthew.James@anu.edu.au}

\author{H.~I.~Nurdin }
\email{Hendra.Nurdin@anu.edu.au}
\affiliation{Department of Engineering, Australian National University, Canberra, ACT 0200, Australia.}

\date{\today}% It is always \today, today,

\begin{abstract}
The aim of this paper is to extend linear quantum dynamical network theory to include static Bogoliubov components (such as squeezers).
Within this integrated quantum network theory   we  provide general methods for cascade or series connections, as well as feedback interconnections using linear fractional transformations.  
In addition, we  define input-output maps and transfer functions for representing components and describing convergence. 
We also discuss the underlying group structure in this theory arising from series interconnection. Several examples illustrate the theory.

\end{abstract}

\maketitle

\section{Introduction}
\label{sec:intro}

The aim of this paper is to develop systematic methods for describing and manipulating a class of quantum feedback networks.
 Quantum networks are important because of the fundamental role they play in quantum technology, and quantum information, computing, and control in particular.  Knowledge of how to interconnect quantum (and classical) components in a network is a pre-requisite for feedback control.

A {\em quantum feedback network} (QFN) \cite{GJ08a}, \cite{GJ08} is a physical network whose nodes (components) are open quantum systems and whose branches (wires) are quantum fields. 
Quantum dynamical components have linear relations between their input and output fields, but in general  their internal physical variables may evolve nonlinearly.  Furthermore, delays may be present due to the non-zero physical length of the  branches and the finite speed of light. The network theory developed in \cite{GJ08a}, \cite{GJ08} is based on quantum stochastic models of open quantum systems (Hudson-Parthasarathy \cite{HP84}, Gardiner-Collett \cite{GC85}), and
provides methods and tools for QFN  modeling including series (or cascade) connections, and feedback loops. The series connection defines a group operation in the class of open quantum dynamical systems. In this paper we will consider the subclass of networks consisting of dynamical components whose internal variables evolve linearly. In \cite{GGY08} the theory was presented for components based on unitary transformations such as beamsplitters and phase shift modulators, but does not include static components that require an external source of quanta for their operation as such amplifiers and squeezers.

In quantum optics one encounters  the class of quantum linear networks with components implementing static linear transformations, called {\em Bogoliubov transformations}, see \cite{Shale}.  Examples of Bogoliubov components include devices capable of creating squeezed states of field (\lq\lq{squeezers}\rq\rq) out of a vacuum, a non-unitary process.
Series connection of static Bogoluibov components is given by matrix multiplication of representations of the Bogoliubov transformations, and defines a group product for this class of systems.

These two classes of quantum networks, which are characterized by the nature of their components, are distinct, but have elements in common. The beamsplitter and phase shift are components in both classes. However, the squeezer does not belong to the class of open dynamical quantum systems (in the framework of Hudson-Parthasarathy \cite{HP84} and Gardiner-Collett \cite{GC85}), though it may be approximated by systems that are in this class.

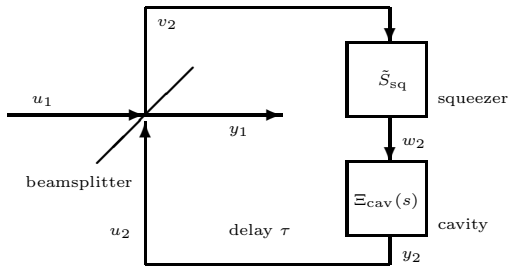
\begin{figure}[h]
\begin{center}

\setlength{\unitlength}{1973sp}%
\begingroup\makeatletter\ifx\SetFigFont\undefined%
\gdef\SetFigFont#1#2#3#4#5{%
 \reset@font\fontsize{#1}{#2pt}%
 \fontfamily{#3}\fontseries{#4}\fontshape{#5}%
 \selectfont}%
\fi\endgroup%
\begin{picture}(5437,3269)(579,-4658)
\put(5251,-2386){\makebox(0,0)[lb]{\smash{{\SetFigFont{6}{7.2}{\familydefault}{\mddefault}{\updefault}{\color[rgb]{0,0,0}$\tilde S_\text{sq}$}%
}}}}
\thicklines
{\color[rgb]{0,0,0}\put(601,-2761){\vector( 1, 0){1725}}
}%
{\color[rgb]{0,0,0}\put(2326,-2761){\vector( 1, 0){1725}}
}%
{\color[rgb]{0,0,0}\put(4876,-4261){\framebox(975,900){}}
}%
{\color[rgb]{0,0,0}\put(2326,-2761){\line( 0, 1){1350}}
\put(2326,-1411){\line( 1, 0){3075}}
\put(5401,-1411){\vector( 0,-1){450}}
}%
{\color[rgb]{0,0,0}\put(5401,-2761){\vector( 0,-1){600}}
}%
{\color[rgb]{0,0,0}\put(5401,-4261){\line( 0,-1){375}}
\put(5401,-4636){\line(-1, 0){3075}}
\put(2326,-4636){\vector( 0, 1){1800}}
}%
{\color[rgb]{0,0,0}\put(4876,-2761){\framebox(975,900){}}
}%
\put(901,-2611){\makebox(0,0)[lb]{\smash{{\SetFigFont{6}{7.2}{\familydefault}{\mddefault}{\updefault}{\color[rgb]{0,0,0}$u_1$}%
}}}}
\put(3376,-2986){\makebox(0,0)[lb]{\smash{{\SetFigFont{6}{7.2}{\familydefault}{\mddefault}{\updefault}{\color[rgb]{0,0,0}$y_1$}%
}}}}
\put(2476,-1636){\makebox(0,0)[lb]{\smash{{\SetFigFont{6}{7.2}{\familydefault}{\mddefault}{\updefault}{\color[rgb]{0,0,0}$v_2$}%
}}}}
\put(1876,-4261){\makebox(0,0)[lb]{\smash{{\SetFigFont{6}{7.2}{\familydefault}{\mddefault}{\updefault}{\color[rgb]{0,0,0}$u_2$}%
}}}}
\put(3376,-4261){\makebox(0,0)[lb]{\smash{{\SetFigFont{6}{7.2}{\familydefault}{\mddefault}{\updefault}{\color[rgb]{0,0,0}delay $\tau$}%
}}}}
\put(5551,-4561){\makebox(0,0)[lb]{\smash{{\SetFigFont{6}{7.2}{\familydefault}{\mddefault}{\updefault}{\color[rgb]{0,0,0}$y_2$}%
}}}}
\put(5551,-3136){\makebox(0,0)[lb]{\smash{{\SetFigFont{6}{7.2}{\familydefault}{\mddefault}{\updefault}{\color[rgb]{0,0,0}$w_2$}%
}}}}
\put(4951,-3886){\makebox(0,0)[lb]{\smash{{\SetFigFont{6}{7.2}{\familydefault}{\mddefault}{\updefault}{\color[rgb]{0,0,0}$\Xi_\text{cav}(s)$}%
}}}}
\put(6001,-4186){\makebox(0,0)[lb]{\smash{{\SetFigFont{6}{7.2}{\familydefault}{\mddefault}{\updefault}{\color[rgb]{0,0,0}cavity}%
}}}}
\put(6001,-2611){\makebox(0,0)[lb]{\smash{{\SetFigFont{6}{7.2}{\familydefault}{\mddefault}{\updefault}{\color[rgb]{0,0,0}squeezer}%
}}}}
\put(826,-3661){\makebox(0,0)[lb]{\smash{{\SetFigFont{6}{7.2}{\familydefault}{\mddefault}{\updefault}{\color[rgb]{0,0,0}beamsplitter}%
}}}}
{\color[rgb]{0,0,0}\put(2926,-2161){\line(-1,-1){1200}}
}%
\end{picture}%

\caption{Linear quantum  feedback network (LQFN) consisting of a beamsplitter, squeezer, and  cavity. The time delay around the optical loop is $\tau$.}

\label{fig:bs-dyn}
\end{center}
\end{figure}

The purpose of this paper is  merge together these two classes of linear networks in a unified, {\em multivariable} algebraic framework. By \lq{multivariable}\rq\ we mean that the framework allows for systems comprised of  multiple oscillator modes and multiple field channels; accordingly, a vector-matrix notation is used. The new class of  linear QFNs  (LQFNs) we consider in this paper are therefore assembled from components of the following two types: (i) dynamical components, with linear evolution of physical variables, and (ii) static components characterized by Bogoliubov transformations. An example of such a network is shown in Figure \ref{fig:bs-dyn}, \cite{YK03a}, \cite{GJ08a}.

To this end, we consider  linear open quantum components that are in general a   series connection of a linear dynamical part, and a static Bogoliubov part. We define series connections of these  components, and extend linear fractional transformation (LFT) methods for describing feedback loops. The series connection defines a group structure for this new class of systems, which includes the linear dynamic and static Bogoliubov classes as subgroups. This group structure is interesting from a physical as well as a systems and control theoretic point of view.

Open quantum systems have a natural input-output structure. We define and make use of input-output maps for  our class of linear open quantum systems, and discuss convergence of systems in these terms. This input-output notion of convergence is important for applications, and is weaker than stronger notions of convergence involving all system variables.

We begin in section \ref{eg:bog} by describing Bogoliubov transformations, which is followed in section \ref{sec:components} by a discussion of open linear dynamical models of Hudson-Parthasarathy type. In section \ref{sec:components-bog} we discuss quantum components involving Bogoliubov transformations, both static and dynamic.  Linear quantum feedback networks are described in section \ref{sec:lqfn}. Several examples are discussed in sections \ref{sec:eg} and \ref{sec:eg-2}.

\section{Bogoliubov Transformations}
\label{eg:bog}

In this section we present models for the quantum components considered in this paper. Before this can be done, some notation is needed.

\subsection{Notation}
\label{sec:components-notation}

Let $X=( X_{jk} )$, $j,k =1,\ldots, n$, denote a matrix whose entries $X_{jk}$ are operators on a Hilbert space $\mathfrak H$, or are complex numbers.
We define the matrices
$$
X^\# = ( X_{jk}^\ast ), \ \
X^\top = ( X_{kj} ), \ \
X^\dagger = ( X^\ast_{kj} ).
$$
Here, the asterisk $\ast$ indicates Hilbert space adjoint or complex conjugation.

For a column vector  $x$ of operators of length $k$, we shall
introduced the \textit{doubled-up} \textit{column vector}
\begin{equation}
\breve{x}\triangleq \left[
\begin{array}{c}
x \\
x^{\#}
\end{array}
\right]
\end{equation}
of length $2k$,
so that  $
\breve{x}^{\dag }= ( x^{\dag },x^{\top }) .
$

Given a linear transformation of the form
$$
y=E_- x+ E_+ x^{\#}
$$
where  $x$ and $y$ are vectors of operators of lengths $k$ and $r$
respectively, and $E_\pm \in \mathbb{C}^{ r \times k}$, we define
the transformation $%
y^{\#}=E_-^{\#}x^{\#}+E_+^{\#}x$, and in doubled-up notation we
have
\begin{equation*}
\breve{y}=    \Delta(E_-,E_+)  \,\breve{%
x},
\end{equation*}
where we introduce the $2r \times 2k$ \textit{doubled-up matrix}
\begin{equation}
\Delta(E_-,E_+)  \triangleq \left[
\begin{array}{cc}
E_- & E_+ \\
E_+^{\#} & E_-^{\#}
\end{array}
\right].
\end{equation}
We note that $\Delta(E_-,E_+)^\dagger= \Delta(E_-^\dagger,
E_+^\top)$, and when the dimensions are compatible,
$\Delta(E_{-},E_{+}) \Delta(F_{-},F_{+}) =
\Delta(E_{-}F_{-}
+ E_{+} F_{+}^\#,
E_{-}F_{+}+ E_{+} F_{-}^\#)$. In the examples we consider, the linear transformations will
be between vectors of equal dimensions and so the matrices $E_\pm$, etc., will be square.

For a $2n \times 2m$ matrix $X$, we define an involution $\flat $ by
\begin{equation}
X^{\flat }\triangleq J_{m}X^{\dag }J_{n},
\end{equation}
where
\begin{equation}
J_{n}\triangleq \left[ 
\begin{array}{cc}
I_{n} & 0 \\ 
0 & -I_{n}
\end{array}
\right] ,
\label{J}
\end{equation}
with $I_{n}$ the $n\times n$ identity matrix. When understood, we shall often drop the dimension index and just write $I$ and $J$. For doubled-up matrices, we then
have 
\begin{equation}
\Delta  ( E_{-},E_{+} ) ^{\flat }=\Delta ( E_{-}^{\dag
},-E_{+}^{\top } ) .
\end{equation}

\subsection{Canonical Commutation Relations}
We consider a collection of $m$ harmonic oscillators, whose
behavior  is characterized by independent annihilation $a_j$ and creation $a_j^\ast$ operators ($j=1,\cdots ,m$)
satisfying the canonical commutation relations   $ [ a_{j},a_{k}^{\ast }] =\delta _{jk}$,
with $ [ a_{j},a_{k}] =0= [ a_{j}^{\ast
},a_{k}^{\ast }] $.
The commutation relations may be written compactly as
\begin{equation*}
[ \breve{a}_{j},\breve{a}_{k }^{\#}] =J_{jk },
\end{equation*}
where $(J_{jk})=J_m$ is the matrix defined in (\ref{J}).

Systems consisting of $m$ oscillator modes are equivalent, for fixed $m$, and it is convenient to consider just the category $\mathcal{S}(m)$ of such systems with representative described by column vector $a=(a_1, \ldots, a_m)^\top$.

\subsection{The Bogoliubov Matrix Lie Group, $\mathrm{Sp} ( \mathbb{C}^m) $}
\label{sec:Bog-Lie-group}

\textbf{Definition}
\textit{A  $2m \times 2m$ complex matrix $\tilde{S}$  is said to be {\em $\flat
$-unitary}   if it is invertible and
\begin{equation*}
\tilde{S}^{\flat }\tilde{S}=\tilde{S}\tilde{S}^{\flat }=I_{2m}.
\end{equation*}
The group of {\em Bogoliubov matrices} $\mathrm{Sp} ( \mathbb{C}^m) $ is the subgroup of $%
\flat $-unitary matrices that are of doubled-up form, that is $\tilde{S}=\Delta(S_-,S_+)$  for suitable  $S_-,S_+ \in \mathbb{C}^{m\times m}$. This is also known as the {\em symplectic group} \cite{KB06}.}

%\bigskip

The transformation $a'=S_-a+S_+a^{\#}$ is called a {\em Bogoliubov transformation} for $a\in \mathcal{S}(m)$. In doubled-up notation this takes the simpler form
\begin{equation}
\breve{a}' =  \tilde{S} \breve a .
\label{S-single-def}
\end{equation}
Note that $a' \in \mathcal{S}(m)$ and in particular, the transformation preserves the canonical commutation relations.

A Bogoliubov matrix $\tilde{S} \in \mathrm{Sp} ( \mathbb{C}^m) $ admits a {\em Shale decomposition} \cite{Shale}
\begin{equation}
\tilde{S}= \Delta(S_{\mathrm{out}}^{\dag },0)
\Delta( \cosh R
,\sinh R )
\Delta( S_{\mathrm{in}},0),
\label{S-decomp}
\end{equation} 
where $S_{\mathrm{in}},S_{\mathrm{out}} $ are $m \times m$ unitary matrices, and $ R$
a real diagonal $m \times m$ matrix. Note that $\Delta (\cosh  R ,\sinh  R )
 = \exp \Delta(0, R )$.
The middle term in (\ref{S-decomp}) corresponds to {\em squeezing},  an important characteristic widely exploited in applications of quantum optics. To see what this means, suppose $\tilde{S}=\Delta( \cosh  R ,\sinh  R )$. Define the quadratures  $a^x = \frac{1}{2}(a+a^\#)$ and $a^y=\frac{1}{2i}(a-a^\#)$, and similarly for $a'$. Then
\begin{equation}
(a')^x = e^ R a^x, \ \ (a')^y = e^{- R} a^y,
\label{S-squeeze}
\end{equation}
which shows that if the $y$ quadrature is scaled by less than unity, the $x$ quadrature must correspondingly be expanded by an amount greater than unity. Also, note that the unitary group $U(m)$ of unitary $m\times m$ matrices can be viewed as a subgroup of  $\mathrm{Sp} ( \mathbb{C}^m) $ via the correspondence $\Delta(S,0) \in \mathrm{Sp} ( \mathbb{C}^m) $ whenever $S\in U(m)$.

The Bogoliubov transformation (\ref{S-single-def}) defined by a fixed $\tilde{S} \in \mathrm{Sp}(\mathbb{C}^m)$ corresponds to the action of a physical device acting on a vector $a \in \mathcal{S}(m)$. By Shale's theorem \cite{Shale} the Bogoliubov transformation may be unitarily implemented, that is, there exists a unitary operator $U$ such that
\begin{equation}
\tilde{S} \breve a = U^\ast \breve a U .
\label{S-single-U}
\end{equation}

\subsection{The Bogoliubov Lie Algebra,  $\mathfrak{sp} (
\mathbb{C}^{m}) $}
\label{sec:Bog-Lie-algebra}

We remark that the Lie algebra $\mathfrak{{sp} ( \mathbb{C}^{m}) }$
consists of matrices $-i \tilde{\Omega}\in \mathbb{C}^{2m\times 2m}$ that are of doubled-up
form (in order to generate doubled-up matrices $\tilde{S}=e^{-i \tilde{\Omega}}$) and satisfy
$ \tilde{\Omega}^{\flat }= \tilde{\Omega}$.
The second condition can be written as $J \tilde{\Omega}- \tilde{\Omega}^{\dag }J=0$. We therefore
deduce that the infinitesimal generators take the form
\begin{equation}
-i \tilde{\Omega}=-\Delta  ( i\Omega_- ,i\Omega_+ ) ,
\label{Hamiltonian-Omega}\end{equation}
with complex matrices $\Omega_-$ and $\Omega_+$ having the
symmetries $\Omega_- ^{\dag }=\Omega_- $ and $\Omega_+^{\top
}=\Omega_+$. We remark that we may construct a hermitean operator $H$ on the oscillator Hilbert space as
\begin{equation}
H =\sum_{\alpha,\beta=1}^{m} ( a_{\alpha }^{\ast
}\omega^-_{\alpha \beta }a_{\beta }+\frac{1}{2} a_{\alpha }^{\ast
}a_{\beta }^{\ast }\omega^+_{\alpha \beta }+\frac{1}{2} a_{\alpha
}a_{\beta }\omega^{+\ast }_{\alpha \beta } ),
\label{Hamiltonian}
\end{equation}
where the coefficients are the entries of the  matrices
$\Omega_\pm = ( \omega^\pm_{\alpha \beta } ) \in \mathbb{C}^{m\times m}$.
From the familiar quantum mechanical point of view, $H$ is the Hamiltonian generating 
the canonical transformation (\ref{S-single-def}), that is, in (\ref{S-single-U}) we have $U= e^{-iH}$.

It is instructive to look at the $m=1$ case. Here the Hamiltonian is $H=\omega_- a^\ast a +\frac{1}{2} \omega_+ a^{\ast 2} + \frac{1}{2} \omega_+^\ast a^2$ with $\omega_-$
real and $\omega_+ = | \omega_+ | e^{i\theta}$ complex. The corresponding element $-i \tilde{\Omega}\in $ $\mathfrak{{sp} ( \mathbb{C}) }$ can be written in terms of Pauli matrices as
as 
\begin{eqnarray*}
-i \tilde{\Omega} &=& -\Delta  ( i\omega_- ,i\omega_+) \\
&=& \omega_{+y} \sigma _{x}+\omega_{+x}
\sigma_{y}-i\omega_- \sigma _{z},
\end{eqnarray*}
where $\omega_+=\omega_{+x}+i\omega_{+y}$, or
\begin{equation*}
\tilde{\Omega}=\left[
\begin{array}{cc}
\omega _{-} & \omega _{+x}+i\omega _{+y} \\ 
\omega _{+x}-i\omega _{+y} & -\omega _{-}
\end{array}
\right] .
\end{equation*}
The eigenvalues of $-i\tilde{\Omega}$ are $\pm \sqrt{\zeta }$ where
\begin{equation}
\zeta =|\omega_+|^2-\omega_-^{2}.
\label{zeta}
\end{equation}
We note that Heisenberg dynamical equations will be trigonometric type for $\zeta <0$, and hyperbolic type for $\zeta>0$. 
Let us try and diagonalize the Hamiltonian by introducing the Bogoliubov transformation $ e^{i \theta} a = \cosh r \, a' - \sinh r \, a'^\ast$, (this is the original purpose of Bogoliubov transformations!). For $\zeta <0$ we may choose $\tanh 2r = |\omega_+ | / \omega_-$ to get $H \equiv \omega_- \sqrt{ 1- | \frac{\omega_+ }{\omega_- } |^2} a'^\ast a'$. While for $\zeta <0$ we may choose $\tanh 2r = \omega_-  / | \omega_+ |$ to get $H \equiv \frac{1}{2} |\omega_+ | \sqrt{ 1- | \frac{\omega_- }{\omega_+ } |^2 }(a'^{\ast 2} + a'^2)$. That is, for $\zeta <0$ we may diagnolize $\Omega$ as ${\tilde S}^\flat \Omega \tilde S = \Delta ( \omega'_- ,0)$ using a Bogoliubov matrix $\tilde S$, but for $\zeta >0$ the best we can do is to put into the form $\Delta ( 0, \omega'_+)$.

We say that $H$ is {\em passive} if $\tilde \Omega$ has only real eigenvalues. In this case we may find an $\tilde S$ in $\mathrm{Sp} ( \mathbb{C}) $ such that $ {\tilde S}^\flat \tilde \Omega \tilde S = \Delta ( \Omega_-',0)$ for some $\Omega_-'$.
The term \lq\lq{passive}\rq\rq \ means that such Hamiltonians do not describe energy flow into the system from an external pumping source, and that the dynamical equations are always trigonometric type.

(For $m=1$, $H$ is passive  if and only if the parameter $\zeta\leq 0$ in (\ref{zeta}) as the eigenvalues are $\pm \sqrt{- \zeta }$. )

The group $\mathrm{Sp} ( \mathbb{C}^{m}) $ is a non-compact group and, in fact, is not covered by the
exponential mapping on its Lie algebra $\mathfrak{{sp} ( \mathbb{C}%
^{m}) }$. We see this in the case $m=1$, where the form of $e^{-i \tilde{\Omega}}$ depends on the sign of  $\zeta$. We have $e^{-i \tilde{\Omega}}$ given respectively by ($\zeta >0$)
{\small
\begin{equation*}
\left[
\begin{array}{cc}\cosh \sqrt{\zeta }-i\omega_- \frac{\sinh \sqrt{\zeta }}{\sqrt{\zeta }} &
\omega_+ \frac{\sinh \sqrt{\zeta }}{\sqrt{\zeta }} \\
\omega_+^{\ast }\frac{\sinh \sqrt{\zeta }}{\sqrt{\zeta }} & \cosh \sqrt{%
\zeta }+i\omega_- \frac{\sinh \sqrt{\zeta }}{\sqrt{\zeta }}
\end{array}
\right], 
\end{equation*}}
and ($\zeta <0$)
{\small
\begin{equation*}
\left[
\begin{array}{cc}
\cos \sqrt{-\zeta }-i\omega_- \frac{\sin \sqrt{-\zeta
}}{\sqrt{-\zeta }} &
\omega_+ \frac{\sin \sqrt{-\zeta }}{\sqrt{-\zeta }} \\
\omega_+^{\ast }\frac{\sin \sqrt{-\zeta }}{\sqrt{-\zeta }} & \cos \sqrt{%
-\zeta }+i\omega_- \frac{\sin \sqrt{-\zeta }}{\sqrt{-\zeta }}
\end{array}
\right] .
\end{equation*}
}
Also $e^{-i \tilde{\Omega}} = 1-i \tilde{\Omega}$ if
$\zeta =0$ (note that $ \tilde{\Omega}^2 =0$ in this case). As observed in \cite{LN04}, we must have \textrm{tr}
$e^{-i \tilde{\Omega}}\geq -2
$ so that there exist matrices in $\tilde{S}\in \mathrm{Sp}%
( \mathbb{C}) $ which do not possess a logarithm in  $\mathfrak{{sp}%
( \mathbb{C}) }$, for example, $\tilde{S}=-\Delta (\cosh
u,\sinh u)\equiv \exp( -\Delta  ( i\pi ,0) ) \exp
(-\Delta  ( 0,-u) ) $. In particular, such
Bogoliubov transformations are not generated by a single
Hamiltonian $H$. The best that can be done is to write the unitary $U$ in (\ref{S-single-U}) as $U=U_1 \cdots U_k$ where each $U_i$ has a logarithm in $\mathfrak{{sp} ( \mathbb{C}^{m}) }$, see \cite {LN04} for higher-order cases.

\subsection{Gaussian States}
\label{sec:GAW}

A state on $\mathcal{S}\left( m\right) $ is said to be Gaussian if we have
\begin{equation*}
\left\langle \exp i\left\{ \breve{u}^{\dag }\breve{a}\right\} \right\rangle
=\exp \left\{ -\frac{1}{2}\breve{u}^{\dag }F\breve{u} + i \breve{u}^{\dag }\breve{\alpha}\right\} ,
\end{equation*}
where $F\geq 0$. For simplicity we consider mean zero states ($\alpha =0$). In particular, we have that $F= \left\langle \breve{a}\breve{a}^{\dag }\right\rangle $ takes the specific form
\begin{equation}
F=\left[
\begin{array}{cc}
\left\langle aa^{\dag }\right\rangle  & \left\langle aa^{\top }\right\rangle 
\\ 
\left\langle a^{\#}a^{\dag }\right\rangle  & \left\langle a^{\#}a^{\top
}\right\rangle 
\end{array}
\right] =\left[
\begin{array}{cc}
I+N^{\top } & M \\ 
M^{\dag } & N
\end{array}
\right] 
\label{F}
\end{equation}
with
\begin{equation}
N_{jk}=\langle a_{j}^{\ast }a_{k}\rangle ,\quad
M_{jk}=\left\langle a_{j}a_{k}\right\rangle 
\label{N+M}
\end{equation}
and we note that $N=N^{\dag }$ and $M=M^{\top }$. In particular, positivity
of $F$ implies that $N\geq 0$. The vacuum state is the special state
determined by the choice $N=0,M=0$, for 
\begin{equation}
F_{\text{vac}}=\left[
\begin{array}{cc}
I & 0 \\ 
0 & 0
\end{array}
\right] .
\label{F-vac}
\end{equation}

For fixed $N\geq 0$ the choice of $M$ is constrained by the requirement that 
$F$ be positive. For the $n=1$ case, $N$ and $M$ are scalars and the positivity condition is
easily seen to be $N\geq 0$ with $|M|^{2}\leq N\left( N+1\right) $. More
generally we should have a diagonalization $V^{\dag }NV=\mathrm{diag}\left(
N_{1},\cdots ,N_{n}\right) $ for unitary $V$ in which case we could consider
new fields $a^{\prime }=Va$. Here $N_{j}$ can be interpreted as the average
number of quanta in the mode $a_{j}^{\prime }$. In general we cannot expect
to simultaneously diagonalize $N$ and $M$.

\subsubsection{Generalized Araki-Woods Representation}

Given a Gaussian state determined by $F$ in equation (\ref{F}), we now show that we can
construct modes having that state through canonical transformations of
vacuum modes. That is, given a state for which $a \in \mathcal{S}(m)$ has covariance $F$ given by (\ref{F}), there exists a $2m \times 4m$ matrix $\tilde S_0$ such that
\begin{equation}
\breve a = \tilde S_0 \breve a_0
\label{aw-1}
\end{equation}
where $a_0 =\left[ 
\begin{array}{c}
a_{1} \\ 
a_{2}
\end{array}
\right] \in \mathcal{S}(m+m)$  has vacuum statistics, and
\begin{equation}
 \tilde S_0 \tilde S_0^\flat = I .
\label{aw-2}
\end{equation}
Indeed, we will construct $\tilde S_0 = \Delta( E^0_-, E^0_+)$  for some
 $m \times 2m$
a matrices $E^0_-$, $E^0_+$.
This generalizes a construction originally due to Araki and Woods \cite{ArakiWoods} for non-squeezed thermal states, see \cite{JGGAW,HHKKR02}.

\subsubsection{Construction of Araki-Woods Vacuum Representation}

%\subsubsection{Diagonalize $N$}

{\em Step 1: Diagonalize $N$.}
We may find a unitary matrix $V\in \mathbb{C}^{m\times m}$ such that $%
V^{\dag }NV= \text{diag} \left( N_{1},\cdots ,N_{m}\right) $. The eigenvalues are
assumed to be ordered such that $N_{1}\geq \cdots \geq N_{m}\geq 0$.
Therefore we can restrict our attention to the case where $N$ is
diagonalized in this way.

{\em Step 2: Ignore zero eigenvalues.} 
Take the first $m_{+}$ eigen values to be strictly positive, with the
remaining $m_{0}=m-m_{+}$ to be zero. We respect to the eigen decomposition $%
\mathbb{C}^{m}=\mathbb{C}^{m_{+}}\oplus \mathbb{C}^{m_{0}}$ we decompose $F$
as
\begin{equation*}
F=\left[ 
\begin{array}{cccc}
I+N_{++} & 0 & M_{++} & M_{+0} \\ 
0 & I & M_{0+} & M_{00} \\ 
M_{++}^{\top } & M_{+0}^{\top } & N_{++} & 0 \\ 
M_{0+}^{\top } & M_{00}^{\top } & 0 & 0
\end{array}
\right] .
\end{equation*}

However, we observe that if a positive matrix has a zero on a diagonal, then
every entry on the corresponding row and column must vanish \cite{footnote} so that actually
\begin{equation*}
F\equiv \left[ 
\begin{array}{cccc}
I+N_{++} & 0 & M_{++} & 0 \\ 
0 & I & 0 & 0 \\ 
M_{++}^{\top } & 0 & N_{++} & 0 \\ 
0 & 0 & 0 & 0
\end{array}
\right] .
\end{equation*}
Therefore, we can restrict our attention to the case where $N$ is diagonal and strictly
positive, and in particular invertible.

{\em Step 3: Explicit Construction.}
We begin by noting the constraint $I+N\geq M\frac{1}{N}M^{\dag }$, which follows from noting the positivity of
\begin{equation*}
\left[ 
\begin{array}{cc}
I & -M\frac{1}{N} \\ 
0 & 0
\end{array}
\right] F
\left[ 
\begin{array}{cc}
I & -M\frac{1}{N} \\ 
0 & 0
\end{array}
\right]^\dag =
\left[ 
\begin{array}{cc}
I+N- M\frac{1}{N}M^{\dag } & 0 \\ 
0 & 0
\end{array}
\right].
\end{equation*}
We then introduce the following matrices:
\begin{eqnarray*}
X &=&\sqrt{I+N-M\frac{1}{N}M^{\dag }}, \\
Y &=&\sqrt{N}= \text{diag} \left( \sqrt{N_{1}},\cdots ,\sqrt{N_{m}}\right) , \\
Z &=&MY^{-1}.
\end{eqnarray*}

Note that $Y = Y^\top$ and from $Z = M Y^{-1}$ we have that $Y Z^\top = M^\top =M
=ZY =ZY^\top$. These matrices satisfy the conditions
\begin{equation}
XX^{\dag }-YY^{\dag }+ZZ^{\dag }=I\text{ and }YZ^{\top }=Z Y^{\top }.
\label{aw-3}
\end{equation}
Now take $b_1$ and $b_2$ to be independent
(commuting) modes in $\mathcal{S}\left( m\right) $. We fix the state to be
the joint vacuum state for both of these modes. 
Then we may represent $a$ as
\begin{equation}
a = X b_1 + Y b_2^\sharp + Z b_2;
\label{aw-4}
\end{equation}
indeed, it is straightforward to check that
\begin{eqnarray*}
\left\langle a^{\#}a^{\top }\right\rangle  &=&Y^{2}=N \\
\left\langle aa^{\top }\right\rangle  &=&Z Y^\top = ZY =M.
\end{eqnarray*}

Therefore we have obtained the representation (\ref{aw-1}) with $\tilde S_0 = \Delta( E^0_-, E^0_+)$ and
$$
E^0_- = \left( \begin{array}{cc}
X & 0
\end{array} \right), \ \ 
E^0_+ = \left( \begin{array}{cc}
Z & Y
\end{array} \right).
$$
Property (\ref{aw-2}) follows from (\ref{aw-3}).

\section{Quantum Open Linear Dynamics}
\label{sec:components}

In this section we consider the general class of open linear dynamical models arising from a unitary model for the joint system and field.
The system will be a collection $\mathcal{S}(m)$ of $m$ harmonic modes with representative  $a=(a_1, \ldots, a_m)^\top$.

\subsection{Boson Fields: Vacuum States}
\label{sec:components-fields}

The open quantum systems to be described below are driven by $n$ quantum
noise fields (input processes) represented by annihilation $b_{j}(t)$ and
creation $b_{j}^{\ast }(t)$ operators ($j=1,\cdots ,n$) satisfying canonical
commutation relations $[b_{j}(t),b_{k}^{\ast }(t^{\prime })]=\delta
_{jk}\delta (t-t^{\prime }),$ with $[b_{j}(t),b_{k}(t^{\prime })]=0$. This
may be written compactly as 
\begin{equation}
\lbrack \breve{b}_{j}(t),\breve{b}_{k}^{\#}(t^{\prime })]=J_{jk}\;\delta
(t-t^{\prime }).
\end{equation}

\bigskip

We shall denote the class of $m$ independent input processes by $\mathcal{F}(n)$ with representative described by column vector $b=(b_1, \ldots, b_n)^\top$.

\bigskip

The vacuum state for the field is characterized by 
\begin{eqnarray*}
\left\langle \exp i\int_{0}^{\infty }\left\{ u\left( t\right) b^{\dag
}\left( t\right) +u^{\dag }\left( t\right) b\left( t\right) \right\}
dt\right\rangle _{\text{vac}}\\
= \exp -\frac{1}{2}\int_{0}^{\infty }u^{\dag
}\left( t\right) u\left( t\right) dt.
\end{eqnarray*}
It is convenient to introduce the integrated fields 
\begin{equation*}
B_{j}(t)=\int_{0}^{t}b_{j}(r)dr,
\end{equation*}
and in the vacuum representation their future-pointing (It\={o}) increments
satisfy the quantum It\={o} table
\begin{equation*}
\begin{tabular}{l|ll}
$\times $ & $dB_{k}^{\dag }$ & $dB_{k}$ \\ \hline
$dB_{j}$ & $\delta_{jk} dt$ & 0 \\ 
$dB_{j}^{\dag }$ & 0 & 0
\end{tabular}
.
\end{equation*}
We may write the above more compactly as
\begin{eqnarray*}
\left\langle \exp i\int_{0}^{\infty }\breve{u}^{\dag }\left( t\right) \breve{%
b}\left( t\right) dt\right\rangle _{\text{vac}}=\\
\exp -\frac{1}{2}%
\int_{0}^{\infty }\breve{u}^{\dag }\left( t\right) F_{\text{vac}}\breve{u}%
\left( t\right) dt.
\end{eqnarray*}

The vacuum state is then the Gaussian state for which $\left\langle \breve{b}%
\left( t\right) \breve{b}^{\dag }\left( t'\right) \right\rangle _{\text{vac}%
}=F_{\text{vac}} \delta (t-t' )$ and the It\={o} table may be summarized by
\begin{equation*}
d\breve{B}d\breve{B}^{\dag }=F_{\text{vac}}dt.
\end{equation*}
Here $B=(B_1,B_2,\ldots,B_n)^\top$. In the vacuum case we may also define the counting process 
\begin{equation*}
\Lambda _{jk}(t)=\int_{0}^{t}b_{j}^{\ast }(r)b_{k}(r)dr,
\end{equation*}
which may be included in the It\={o} table \cite{HP84}. The additional
non-trivial products of differentials are
\begin{equation*}
d\Lambda _{jk}dB_{l}^{\dag }=\delta _{kl}dB_{j}^{\dag },dB_{j}d\Lambda
_{kl}=\delta _{jk}dB_{l},d\Lambda _{jk}d\Lambda _{li}=\delta _{kl}d\Lambda
_{ji}.
\end{equation*}

\subsection{Boson Fields: Gaussian Field States}
 \label{sec:components-bog-aw}

We may generalize the situation in \ref{sec:components-fields} to the case where the input fields are in Gaussian states with zero mean, but with the
correlation functions
\begin{eqnarray*}
\langle b_{j}^{\ast }(t)b_{k}(t^{\prime })\rangle =N_{jk}\;\delta
(t-t^{\prime }),\\
\langle b_{j}(t)b_{k}(t^{\prime })\rangle =M_{jk}\;\delta
(t-t^{\prime }),
\end{eqnarray*}
with $N$ and $M$ as in (\ref{N+M}). That is, 
\begin{equation}
\left\langle \breve{b}\left( t\right) \breve{b}^{\dag }\left( t^{\prime
}\right) \right\rangle \equiv F\delta \left( t-t^{\prime }\right) ,
\end{equation}
where $F$ has the same form encountered in the case of a finite number of modes in equation (\ref{F}).
The extended It\={o} table is then
\begin{equation*}
\begin{tabular}{l|ll}
$\times $ & $dB_{k}^{\dag }$ & $dB_{k}$ \\ \hline
$dB_{j}$ & $(\delta_{jk} +N_{kj}) dt$ & $M_{jk}dt$ \\ 
$dB_{j}^{\dag }$ & $M_{kj}^\ast dt $ & $ N_{jk} dt$.
\end{tabular}
\end{equation*}

Generalized Araki-Woods representations for arbitrary Gaussian field states may be obtained 
based on a straightforward lifting of the constructions in \ref{sec:GAW}, \cite{JGGAW,HHKKR02}.

\subsection{Quantum Linear Dynamical Models}
\label{sec:components-dyn-models}

The dynamical behavior of a  system comprised of $m$ oscillators,
interacting with $n$ input fields (vacuum state) is given in terms of the Hudson-Parthasarathy 
Schr\"{o}dinger equation (in It\={o} form) \cite{HP84}
\begin{eqnarray}
dU ( t) =\biggl\{\sum_{i,j=1}^{n} ( S_{ij}-\delta
_{ij}) \,d\Lambda _{ij}  ( t) +\sum_{j=1}^{n}
dB_{j}^{\ast } ( t) \,L_{j}   \notag\\
-\sum_{j,k=1}^{n} L_{j}^{\ast
} S_{jk}\,dB_{k} (
t) - 
(\frac{1}{2}\sum_{j=1}^n L_{j}^{\ast
}L_{j}+iH)\,dt\biggr\}U ( t) ,\label{HP unitary QSDE}
\end{eqnarray}
for a unitary operator $U(t)$, with $U(0)=I$. To obtain a unitary evolution leading to linear dynamics we must take $S\in \mathbb{C}^{n\times n}$  to be unitary,  $H$ to be of the form encountered in (\ref{Hamiltonian}), while the coupling of the system modes to the fields is to be of the form
\begin{equation}
L_{j} =\sum_{\alpha=1}^{m} (C^-_{j\alpha }a_{\alpha }+ C^+_{j\alpha }a_{\alpha }^{\ast }), 
\end{equation}
where  $C_\pm =(C^\pm_{j\alpha}) \in \mathbb{C}^{n\times m}$.

The oscillator variables evolve unitarily $a_j(t) = U^\ast(t) a_j
U(t)$, and likewise the output field is $B_{\mathrm{out}}(t)=
U^\ast(t) B(t) U(t)$. The dynamical equations are
\begin{eqnarray}
&&\dot a ( t) =C_+^{\top }S^{\#}\,b^{\#} ( t)
-C_-^{\dag }S\,b ( t) +  A_-a ( t)  + A_+a^{\#} ( t),
\nonumber
\\
&& b_{\mathrm{out}}  ( t) =S\,b ( t) +C_-
a( t ) + C_+ a^{\#}( t ) , \label{lin-dyn-s}%\nonumber
\end{eqnarray}
where
\begin{equation}
A_\mp =-\frac{1}{2} ( C_-^{\dag }C_\mp - C_+^{\top }C_\pm^{\#})
-i\Omega_\mp  .
\end{equation}
Note that $-\frac{1}{2i}(A_- - A_-^{\dag })=\Omega_- $ and
$-\frac{1}{2i}(A_+ + A_+^{\top })= \Omega_+$, but in general
$A_-\neq A_-^{\dag }$ and $ A_+ \neq A_+^{\top }$. Here and below differential 
equations are expressed in terms of quantum noise fields, and may be interpreted in the 
Stratonovich or It\={o} sense, moreover the evolution
preserves the commutation relations of the oscillator variables.

The linear dynamical equations can be written in doubled-up form as
\begin{eqnarray}
\dot{\breve{a}}( t ) &=& \Delta ( A_- ,A_+ ) \breve{a}( t ) -\Delta ( C_- , C_+ )^{\flat } \Delta ( S,0)   \breve{b}  ( t ) ,
\nonumber
\\
\breve{b}_{\mathrm{out}}( t ) &=&\Delta ( C_-,C_+
) \, \breve{a}( t ) +\Delta ( S,0)
\,\breve{b}( t ) . \label{lin-dyn-d1}
\end{eqnarray}

Let us introduce the doubled-up matrices  $\tilde{A}=\Delta ( A_-, A_+ ) $,
$\tilde{C}=\Delta ( C_-, C_+ ) $, and $-i \tilde{\Omega}=-\Delta ( i\Omega_- ,i\Omega_+ )
$, then we have the identities
\begin{equation}
2\mathrm{Re}_{\flat}(A)=\tilde{A}+\tilde{A}^\flat = -\tilde{C}^\flat \tilde{C}, \quad
\tilde{\Omega}^\flat =\tilde{\Omega}.
\label{Q}
\end{equation}

These are readily established by noting
\begin{eqnarray*}
& & \Delta ( A_-,A_+ ) +\Delta ( A_-,A_+ )^{\flat }=\Delta ( A_- + A_-^{\dag }, A_+ - A_+^{\top } )\\
&=&- \Delta ( C_-^{\dag }C_- -C_+^{\top }
C_+^{\#}, C_-^{\dag } C_+ - C_+^{\top}C_-^{\#} )\\
&=& - \Delta ( C_-, C_+  )^{\flat } \Delta ( C_-,C_+) ,
\end{eqnarray*}
and
\begin{equation*}
\Delta ( i\Omega_- ,i \Omega_+ )^{\flat
}=\Delta ( -i\Omega_-^{\dag },-i \Omega_+^{\top } )
=-\Delta( i\Omega_- ,i \Omega_+ ) .
\end{equation*}

The dynamical equations can then be recast as
\begin{eqnarray}
\dot{\breve{a}}( t ) &=&\tilde{A}\breve{a} (
t) +\tilde{B} \breve{b}( t ) , \nonumber
\\
\breve{b}_{\mathrm{ out}}( t )
&=&\tilde{C}\breve{a}( t ) +\tilde{D}\breve{b}  (
t) , \label{lin-dyn-d2}
\end{eqnarray}
where 
\begin{equation}
\tilde{D}=\Delta( S,0) ,
\label{D-S}
\end{equation}
and
\begin{equation}
\tilde{B}=-\tilde{C}\,^{\flat }\tilde{D},\quad
\tilde{A}=-\frac{1}{2}\tilde{C}\,^{\flat }\tilde{C}-i \tilde{\Omega}.
\label{AB}
\end{equation}

We denote this  class of linear Hudson-Parthasarathy systems with $n$ input fields and $m$ oscillators by $\mathcal{L}^{\text{HP}}(n,m)$, and write $\mathcal{L}^{\text{HP}}(n) = \cup_m \mathcal{L}^{\text{HP}}(n,m)$. Systems $G \in \mathcal{L}^{\text{HP}}(n,m)$ may be parameterized in several ways. In terms of the scattering matrix, $S$, vector of coupling operators $L$ and Hamiltonian $H$ we may write
\begin{equation}
G = (S, L, H) .
\label{G-params-0}
\end{equation}
Since these physical parameters are determined by the matrices given above, we may also write
\begin{equation}
G = (S, \tilde C, \tilde \Omega) .
\label{G-params-1}
\end{equation}
Alternatively, we may use  the matrices appearing in equations (\ref{lin-dyn-d2}), 
  \begin{equation}
 G= \left[  \begin{array}{c|c} \tilde A  &   \tilde B
\\
\hline \tilde C & \tilde D
\end{array}
\right]  , \label{open-G-params-d}
\end{equation}
a notation commonly used in linear systems and control theory.
We remark that an arbitrary quadruple of matrices $\tilde A$,
$\tilde B$, $\tilde C$, $\tilde D$ need not necessarily correspond
to a quantum open system, \cite{JNP08}, \cite{NJD08}.

\subsection{Stability}
\label{sec:components-dyn-stab}

In linear systems theory, a system $G$ of the form (\ref{open-G-params-d}) is said to be  {\em Hurwitz stable} if the matrix $\tilde{A}$ has all
eigenvalues having strictly negative real part. Now if
 the joint system-field
state is one in which the inputs are mean-zero, then $\frac{d}{dt}\langle 
\breve{a}\left( t\right) \rangle =\tilde{A}\langle \breve{a}\left( t\right)
\rangle $ and so Hurwitz stability implies that $\langle \breve{a}\left(
t\right) \rangle \rightarrow 0$ as $t\rightarrow \infty $.

When $A_+ =0$, we have $\tilde{A}=\Delta \left( A_{-},0\right) $
with $A_{-}\equiv -\frac{1}{2}C_{-}^{\dag }C_{-}-i\Omega _{-}$ and $\Omega
_{-}^{\ast }=\Omega _{-}$. Since $X^{\dag }X$ is non-negative definite it is
easy to determine whether $A_{-}$ is Hurwitz. For instance, it is sufficient
to have $C_{-}$ invertible.
However, expressions such as $X^{\flat }X$ are indefinite due to the
presence of the matrix $J$. There may be non-passive contributions to $%
\tilde{A}$ from both $C_{+}$ and $\Omega _{+}$. 

As an illustration let us consider  how the eigenvalues of $\tilde A$ depend on the physical parameters in the simplest case $n=1=m$. 
We have seen that the most general parameterization is 
\begin{equation*}
\tilde{C}=\Delta \left( \sqrt{\gamma _{-}}e^{i\phi _{-}},\sqrt{\gamma _{+}}%
e^{i\phi _{+}}\right) \text{ and }\tilde{\Omega}=\Delta \left( \omega
_{-},\omega _{+}\right) ,
\end{equation*}
with $\gamma _{\pm },\phi _{\pm }$ and $\omega _{-}$ real and $\omega
_{+}\in \mathbb{C}$. In this case 
\begin{equation*}
\tilde{A}=-\Delta (\frac{1}{2}\left( \gamma _{-}-\gamma _{+}\right) +i\omega
_{-},i\omega _{+})
\end{equation*}
which has eigenvalues $\frac{1}{2}\left( \gamma _{-}-\gamma _{+}\right) \pm 
\sqrt{\zeta }$ where we recall the parameter $\zeta =|\omega
_{+}|^{2}-\omega _{-}^{2}$ from (\ref{zeta}). The plant is Hurwitz if

\begin{enumerate}
\item  $\zeta \leq 0$ and $\gamma _{-}>\gamma _{+};$

\item  or $\zeta >0$ and $\sqrt{\zeta}<\frac{1}{2}\left( \gamma _{-}-\gamma _{+}\right) $.
\end{enumerate}

In situation 1 the system has a passive Hamiltonian and the damping rate is greater that the pumping rate.  
However situation 2 shows that if the damping is sufficiently large then the system may still be stable even if the Hamiltonian is not passive.  In general, as one expects, stability will depend on the  relative flows of energy into and out of the system.

 \subsection{Series Connections}
 \label{sec:components-dyn-series}

Open linear dynamical systems $G_1=(S_1, L_1, H_1)$ and $G_2=(S_2, L_2, H_2)$ in $\mathcal{L}^{\text{HP}}(n)$
(recall the parameterization (\ref{G-params-0})) may be connected in series by passing the output of system $G_1$ into the input of  system $G_2$, \cite{CWG93}, \cite{HJC93}, \cite{GJ08}. The system formed from this connection in the zero-delay limit is an open system $G=G_2 \triangleleft G_1$, which in terms of
the parameters (\ref{G-params-0}) is given by
\begin{equation}
G_2 \triangleleft G_1 = ( S_2 S_1, L_2+ S_2 L_1, H_1+H_2 + \mathrm{Im} \{  L_2^\dagger S_2 L_1 \} ) .
\label{HP-series-def}
\end{equation}
We refer to $\triangleleft$ as the \textit{series product} of $G_1$ and $G_2$ and gives the cascaded system. The set  $\mathcal{L}^{\text{HP}}(n)$ forms a group with respect to the series product, with inverse $G^{-1}=(S^\dagger, -S^\dagger L, -H)$ (where $G=(S,L,H)$).

Given an open system $G=(S,\tilde C,\tilde \Omega)$ (now we use  the parameterization (\ref{G-params-0}) in anticipation of later use), it follows from properties of the series product that
 \begin{equation}
 G = (I, \tilde C, \tilde \Omega) \triangleleft (S,0,0) .
 \label{G-S-factor}
 \end{equation}
 This factorization says that an open system with scattering $S$ is equivalent (in the zero-delay limit) to a dynamic open system without scattering $(I,\tilde C,\tilde \Omega)$ connected in series with a non-dynamic or static open system $(S,0,0)$.

 \subsection{Input-Output Maps}
 \label{sec:components-dyn-iomaps}

   In classical systems and control theory \cite{ZDG96}  the input-output map is a basic tool, which in the case of linear systems may be expressed explicitly in the time and frequency domains. Input-output maps  for the open quantum linear system may be defined in the same way; in terms of the doubled-up parameters (\ref{open-G-params-d}), we have
\begin{equation}
\breve b_{\mathrm{out}}(t) =  \tilde C e^{\tilde At} \breve a(0) +  \tilde \Sigma_G[t;  \breve b_{\text{in}} ]
\label{G-io-t}
\end{equation}
where
 \begin{equation}
\tilde \Sigma_G[t;  \breve b] =  - \int_0^t \tilde Ce^{\tilde A(t-r)} \tilde C^\flat \tilde D \breve b(r) dr + \tilde D \breve b(t) .
\label{G-io-t-G}
\end{equation}
Here the input  $b_{\text in} $ is understood as the input field $b$ and we shall often use the subscript for emphasis.
 The impulse response associated with the term $ \tilde \Sigma_G[t;  \breve b]$ is $\tilde \sigma_G(t)=-\tilde C e^{\tilde At} \tilde C^\flat \tilde D+\tilde D\delta(t)$, from which we have the transfer function (the Laplace transform of  $ \tilde \Sigma_G[t;  \breve b]$, in which $s$ is a complex variable):
 \begin{equation}
\tilde \Xi_G(s) =  \left[  \begin{array}{c|c}
\tilde A  &  -\tilde C^\flat \tilde D
\\
\hline
\tilde C & \tilde D
\end{array}
\right] (s)= -\tilde C(sI-\tilde A)^{-1}\tilde C^\flat \tilde D +\tilde D.
\label{G-io-s}
\end{equation}

Let us introduce the transformed
\begin{equation} \breve b_{\text{in}}\left[ s\right] \triangleq \int_{0}^{\infty }e^{-st} \breve b_{\text{in}}\left(
t\right) dt ,
\label{bsquare}
\end{equation}
that is, $b_{\text{in}}\left[ s\right] = \int_{0}^{\infty }e^{-st}  b_{\text{in}}\left(
t\right) dt $ and $b_{\text{in}}^{\# }\left[ s\right] =  b_{\text{in}}[s^\ast]^\# 
=\int_{0}^{\infty }e^{-st}b_{\text{in}}^{\#}\left( t\right) dt$. 

Adopting a similar convention for the outputs, we then obtain an input-output
relation of the form $b_{\text{out}} \left[ s\right] = \Xi_{G,-}\left( s\right) b_{\text{in}}\left[ s%
\right] + \Xi_{G,+} \left( s\right) b_{\text{in}}^\ast \left[ s\right]$, or
\begin{equation}
\breve b_{\text{out}}\left[ s\right] = \tilde \Xi_G (s) \breve b_{\text{in}}\left[ s\right] 
\label{IOTF}
\end{equation}
where 
\begin{equation}
\tilde \Xi_G( s) = \left[ 
\begin{array}{cc}
\Xi _{G,-}\left( s\right)  & \Xi _{G,+}\left( s\right) \\ 
\Xi _{G,+}\left( s^{\ast }\right) ^{\#} & \Xi _{G,-}\left( s^{\ast }\right) ^{\#}
\end{array}
\right]
, \label{G-gi-a}
\end{equation}
and we have ignored the initial value contribution of the system modes.

Note that while the transfer function $\tilde \Xi_G(s)$ is uniquely determined by $G$, the transfer function does not uniquely determine the system $G$ - many systems may have the same transfer function.

\section{Examples}
\label{sec:eg}

\subsection{Annihilation Systems}
\label{sec:components-dyn-gauge}

A system $\tilde G = (S, \tilde C, \tilde \Omega)$ with $C_+=0$, $\Omega_+=0$
has dynamics and output relations that depend only on the annihilation operators and annihilation fields,  \cite{GGY08,MP09,Nurd09}. For a physically motivated reason, 
since neither the Hamiltonian (passive) nor the coupling operator of the system contain terms that would require an external source of quanta
(i.e, a classical pump beam) to implement (this follows from the synthesis theory of \cite{NJD08}; see also \cite[Section 7]{Nurd09} for a discussion), they are also referred to as passive systems \cite{Nurd09}. This type of system often arises in applications, and includes optical cavities.
Transfer functions for this class of systems takes a simpler form, as we now describe.

We have    $A_- \equiv -\frac{1}{2} C_-^{\dag }C_- + i\Omega_- $
and $A_+=0$. Then the matrices $\tilde{C}=\Delta (
C_-,0) $ and $\tilde{A}=\Delta( A_-,0) $ are
block diagonal and the transfer function takes the form
\begin{equation}
\tilde \Xi_G( s) = \left[ 
\begin{array}{cc}
\Xi _{G,-}\left( s\right)  & 0 \\ 
0 & \Xi _{G,-}\left( s^{\ast }\right) ^{\#}
\end{array}
\right]
, \label{G-gi-1}
\end{equation}
with
\begin{equation}
\Xi_{G,-}(s) =\left[
\begin{tabular}{l|l}
$A_-$ & $-C_-^{\dag }S$ \\ \hline
$C_-$ & $S$%
\end{tabular}
\right] (s) = -C_- (sI-A_-)^{-1} C_-^\dagger S + S . 
\label{G-gi-2}
\end{equation}
In this situation, we have the input-output relation $ b_{\mathrm{out}}(t) =   C e^{ At}  a(0) +    \Sigma_G[t;   b]$ with
$\Sigma_G[t;   b] =  - \int_0^t  Ce^{ A(t-r)}  C^\flat  D b(r) dr +  D b(t)$. In comparison with (\ref{G-io-t}) and (\ref{G-io-t-G}), the output field depends affinely on $b$, but not the conjugate $b^\sharp$.

\subsection{Cavity}
\label{sec:eg-2-cavity}

In a rotating reference frame, a model for a detuned cavity  is characterized by the parameters $G_\text{cav} = (1, \sqrt{\gamma} a, \omega a^\ast a)$; i.e. $\Omega_-=\omega$, $\Omega_+=0$, $C_-=\sqrt{\gamma}$, $C_+=0$, $S=I$.  

This corresponds to an annihilation-form system
\begin{eqnarray}
\dot a &=& -(\frac{\gamma}{2} +i\omega) a - \sqrt{\gamma} b_\text{in}
\nonumber 
\\
b_\text{out} &=& \sqrt{\gamma} \, a + b_\text{in}
\label{cavity-1}
\end{eqnarray}
when driven by vacuum input $b$.
The transfer function for this system may readily be computed to be
\begin{equation}
\Xi_\text{cav,-}(s) = \frac{s-\frac{\gamma}{2} +i \omega }{s+\frac{\gamma }{2}+ i \omega} ,
\label{cavity-2}
\end{equation}
which in doubled-up form  is
\begin{equation}
\tilde \Xi_\text{cav}(s)  = 
\left[ \begin{array}{cc}
\frac{s-\frac{\gamma}{2} +i \omega }{s+\frac{\gamma}{2} +i \omega} & 0
\\
0 & \frac{s-\frac{\gamma}{2} - i \omega  }{s+\frac{\gamma}{2} -i \omega} 
\end{array} \right].
\label{cavity-3}
\end{equation}
Thus this system is
$$
G_\text{cav} = (I, \Delta( \sqrt{\gamma}\, I, 0), -i \Delta(i \omega , 0))  \in \mathcal{L}^{\text{HP}}(1).
$$

\subsection{Degenerate Parametric Amplifier}
\label{sec:eg-amp}

We consider the model for a degenerate parametric amplifier (DPA)
\cite[sec. 7.2]{GZ00},  which corresponds to a single oscillator
$G$ coupled to a single field with $S=I$,
$\omega_-=0$, $ \omega_+ = \frac{i}{2}\epsilon$,  $\epsilon >0$,
$C_-=\sqrt{\kappa}$, and $C_+=0$. The Hamiltonian will not be passive,  however, the system is stable in the sense of Hurwitz if take $ \epsilon \leq \kappa$ as we shall do from now on.
Using  (\ref{G-io-s}) we find that the doubled-up transfer function, in agreement with \cite{GZ00}, is
\begin{equation*}
\tilde{\Xi}_{\mathrm{DPA}}(s)=\frac{1}{P(s)}\left[ 
\begin{array}{cc}
s^{2}-\frac{\kappa ^{2}+\epsilon ^{2}}{4} & -\frac{1}{2}\epsilon \kappa  \\ 
-\frac{1}{2}\epsilon \kappa  & s^{2}-\frac{\kappa ^{2}+\epsilon ^{2}}{4}
\end{array}
\right] ,
\end{equation*}
where $P(s)= ( s+\frac{1}{2}\kappa )^2 -\frac{1 }{4}\epsilon ^{2}.$ The poles of the transfer function therefore occur at the zeros of $P$, namely $s=\pm \frac{\epsilon }{ 2} - \frac{\kappa }{ 2}$.
In the frequency domain,  the output field is
\begin{eqnarray*}
b_{\mathrm{out}}(s)
= \frac{1}{P(s)}( s^{2}-\frac{\kappa ^{2}+\epsilon ^{2}}{4})
b(s)    -\frac{1}{2P(s)} \epsilon \kappa
b^\ast(s)
.
\end{eqnarray*}
(Here we ignore the initial condition contribution which is justified by the stability of the system.) In terms of quadratures $b^x=\frac{1}{2}(b+b^\ast)$ and $b^y=\frac{1}{2i}(b-b^\ast)$, we find that
$$
b^x_{\mathrm{out}}(s) = \Xi^x_{\mathrm{DPA}} (s) b^x(s), \ \ b^y_{\mathrm{out}}(s)=\Xi^y_{\mathrm{DPA}} (s) b^y(s)
$$
where (in agreement with \cite[eq. (7.2.26)]{GZ00})
\begin{equation*}
\Xi_{\mathrm{DPA}}^{x}( s) =\frac{s-\dfrac{\kappa +\epsilon }{2}}{s+\dfrac{%
\kappa -\epsilon }{2}}=\frac{1}{\Xi_{\mathrm{DPA}}^{y}( s) }.
\end{equation*}

The DPA can be implemented in a single-ended cavity and a case that is of our main interest in this paper is the idealized one (for a full discussion
see \cite[10.2.1.g]{GZ00}) where 
$\kappa,\epsilon \rightarrow \infty$ (in practice to be taken large) such that the ratio $ \frac{\epsilon }{ \kappa }$ is constant. Rescaling $\kappa = k \kappa_0$ and $\epsilon = k \epsilon_0$ is equivalent to replacing $\kappa $ by $\kappa_0$ and $\epsilon $ by $\epsilon_0$ and rescaling $s$ as $\frac{s}{ k}$:
\begin{equation*}
\tilde \Xi_{\mathrm{DPA}}( s,\kappa = k\kappa_0,\epsilon=k\epsilon_0)= \tilde \Xi_{\mathrm{DPA}}( \frac{s}{k}, \kappa_0,\epsilon_0).
\end{equation*}
The limit $k \to \infty$ is the appropriate limit and here the cavity has an instantaneous response and the internal cavity dynamics are essentially eliminated by adiabatic elimination. This results in $b_{\rm out}(s)$ being given as the following Bogoliubov transformation of the input:
\begin{equation*}
b_{\mathrm b} (s) = - \cosh(r_0) \, b(s) - \sinh(r_0) \, b^\dag (s),
\end{equation*}
where
\begin{equation*}
r_0 = \ln \frac{\kappa_0 + \epsilon_0}{\kappa_0 - \epsilon_0}.
\end{equation*}
The output is then an ideal squeezed white noise process satisfying the quantum It\={o} rule discussed in section \ref{sec:components-fields}, where
\begin{eqnarray*}
N =\sinh^2 r_0 = \frac{4 \kappa_0 \epsilon_0 }{ ( \kappa^2_0 -\epsilon_0^2 )^2} ,\\ 
M=\cosh r_0 \sinh r_0 = \frac{2 \kappa_0 \epsilon_0 ( \kappa_0^2 - \epsilon_0^2)}{ ( \kappa^2_0 -\epsilon_0^2 )^2}.
\end{eqnarray*}
Note here that $M$ and $N$ satisfy the relation $| M |^2=N(N+1)$. 
In this limit the DPA device behaves like a static device that instantaneously outputs a squeezed white noise field from a vacuum white noise field source, and the transfer function has a constant Bogoliubov matrix value across {\em all} frequencies. That is
\begin{eqnarray*}
\tilde \Xi_{\mathrm{DPA\,static}}(s) &=& \lim_{k \to \infty} \tilde \Xi_{\mathrm{DPA}}(\frac{s}{k},\kappa_0,\epsilon_0)\\
&= &  - \Delta ( \cosh r_0, \sinh r_0)
, \forall  s \in \mathbb{C}, 
\end{eqnarray*}
and the quadrature transfer functions are 
$$\Xi_{\mathrm{DPA\,static}}^{x}( s) = -e^{r_0}, \, \Xi_{\mathrm{DPA\,static}}^{y}( s) = -e^{-r_0}.$$

For a DPA device with a sufficiently wide bandwidth, one may approximately model it as a static  DPA device with the ideal characteristics described above. Clearly, $\mathcal{L}^{\text{HP}}(n)$ is not closed with respect to this type of approximation.

\section{Components involving Bogoliubov Transformations}
\label{sec:components-bog}

In section \ref{sec:eg-amp} we obtained the constant transfer function $\tilde \Xi_{\mathrm{DPA\,static}} \in \mathrm{Sp}(\mathbb{C})$ for a static approximation to a DPA. Such static approximations afford useful simplifications, though in reality the DPA is a dynamical physical device. The idealized DPA therefore yields outputs that are a squeezing of the inputs.

Motivated by this, in section \ref{sec:components-bog-comp} we consider Bogoliubov matrices acting on boson fields, thereby extending the class of static components beyond unitary scattering devices. These components will be combined with linear dynamics in section \ref{sec:lqfn-comp} to form a general class of quantum linear systems; such models may be useful when the time scales of the dynamical parts are slower than  the time scales of the systems represented by static Bogoliubov matrices. These components will be combined with linear dynamics in section \ref{sec:lqfn-comp} to form a general class of quantum linear systems; such models may be useful when the time scales of the dynamical parts are slower than  the time scales of the systems represented by static Bogoliubov matrices.

\subsection{Bogoliubov Static Components}
\label{sec:components-bog-comp}

More generally we could consider a static component which performs a Bogoliubov transformation of the input field $b_{\mathrm{in}} \in \mathcal{F}(n)$: 
\begin{equation}
\breve b_{\mathrm{out}} (t) =  \tilde{S} \, \breve b_{\mathrm{in}} (t).
\label{S-field-def}
\end{equation}
where now $\tilde{S} \in \mathrm{Sp}(\mathbb{C}^n)$. This transformation, of course, preserves the canonical commutation relations so that $b_{\mathrm{out}}  \in \mathcal{F}(n)$. 

Some caution should be applied here as we are now using the symbol $\tilde{S}$ in (\ref{S-field-def}) in a purely algebraic manner as an element of $\mathrm{Sp}(\mathbb{C}^n)$ when we strictly mean the second-quantization of the Bogoliubov matrix as an operator on the fields. Despite its formal similarity to (\ref{S-single-def}), the relation (\ref{S-field-def}) is of a different character as the fields carry a continuous time-variable. Moreover, since such a transformation in general form linear combinations of field annihilation operator and creation operators, the transformation $\breve b_{\rm out}(t) = \tilde S \breve b_{\rm in}(t)$ cannot be described by the usual Hudson-Parthasarathy QSDE for open Markov systems (cf. Section \ref{sec:components-dyn-models}). Such a QSDE can only model linear combinations of field annihilation operators  of the form $\breve b_{\rm out}(t) = \Delta(S,0) \breve b_{\rm in}(t)$ for a unitary matrix $S$ that appears as one of the parameters of the QSDE (here we set the other parameters to $L=0$ and $H=0$). As such, in the transformation of fields with a non-unitary Bogoliubov matrix we do not have an analogue of (\ref{S-single-U}) in the form of $\breve b_{\rm out}(t) = U(t)^*\breve b_{\rm in}(t) U(t)$ for some unitary process $U(t)$ on the system and noise Hilbert space. At present we do not know whether a unitary transformation exists and if it exists what kind of dynamical equations it would satisfy. Since unitary evolution is a fundamental postulate of quantum mechanics, the situation is  somewhat unsatisfactory and is the subject of continuing research. However, the relation (\ref{S-field-def}) is nevertheless a useful idealization for certain devices used in quantum optics, such as what we have seen with the static DPA in Section \ref{sec:eg-amp}, and has formally been employed up to now (see, for instance, the discussion in Chapter 7 of \cite{GZ00} on various quantum optical amplifiers). The physical meaning of the Bogoliubov transformation (\ref{S-field-def}) is correctly interpreted as a limiting situation.

\subsection{Bogoliubov Static Components as Limits of Dynamical Components}
\label{sec:components-bog-lim}

The class of linear dynamical  components described in section \ref{sec:components-dyn-models} is not closed under input-output convergence. We now show how arbitrary static Bogoliubov components may arise as limits of unitary models. The idea is to exploit the Shale decomposition (\ref{S-decomp}). Thus any given Bogoliubov matrix $\tilde S$ has the decomposition  $\tilde S=\Delta(\tilde S_{\rm out}^{\dag},0) \Delta(\cosh R,\sinh R) \Delta(\tilde S_{\rm in},0)$, where $\tilde S_{\rm in}$ and $\tilde S_{\rm out}$ are some unitary matrices and $R$ is some real diagonal matrix. 
We note that the end terms $\Delta(\tilde S_{\rm out}^{\dag},0)$  and  $\Delta(\tilde S_{\rm in},0)$ can each be realized as a static passive network made of beamsplitters, mirrors and phase shifters. The middle term of course describes squeezing but this arises from a straightforward construction involving $n$ independent static-limit DPAs acting as ideal squeezing devices. (Each DPA corresponding to a diagonal entry of $R$ providing  a degree of squeezing (cf. Section \ref{sec:eg-amp}) as determined by that entry.) Then we note that we may approximate each DPA with a corresponding dynamic (non-ideal) DPA with appropriate parameters (see the discussion of the DPA in Subsection \ref{sec:eg-amp}). 

\subsection{Dynamical Bogoliubov Components}
\label{sec:lqfn-comp}

We introduce an extension of the class of dynamical linear models $\mathcal{L}^{\text{HP}}(n)$ considered
up to now to accommodate the notion of squeezing. This extension is inspired by the factorization (\ref{G-S-factor})  
for open linear systems of $\mathcal{L}^{\text{HP}}\left( n\right) $ type,  suggesting that we consider a new class of dynamical components of the form
\begin{equation}
G = (\tilde{S}, \tilde C, \tilde \Omega)  \triangleq  (I,\tilde C,\tilde \Omega) \triangleleft \tilde{S} ,
\label{G-S-def-0}
\end{equation}
where $(I,\tilde C,\tilde \Omega) \in \mathcal{L}^{\text{HP}}\left( n,m\right) $ and $\tilde{S}=\Delta(S_-, S_+)  \in \mathrm{Sp}(\mathbb{C}^n)$.  

A system $G = (\tilde{S}, \tilde C, \tilde \Omega) $ is defined by  equations  (\ref{lin-dyn-d2}),  where $\tilde A$, $\tilde B$ and $\tilde C$ are as before (section \ref{sec:components-dyn-models}), but now $\tilde D = \tilde S$. 
We use the notation $\mathcal{L}^{\text{Bog.}}(n,m)$ to denote this class of systems and write $\mathcal{L}^{\text{Bog.}}(n) = \cup_m\mathcal{L}^{\text{Bog.}}(n,m)$. The class $\mathcal{L}^{\text{Bog.}}(n) $  includes $\mathcal{L}^{\text{HP}}(n)$ as a special case (with $\tilde S=\Delta(S,0)$).
The justification for  the cascade expression (\ref{G-S-def-0}) will be given in  section \ref{sec:gen-nw-series}, where we shall extend the series product for cascaded systems in $\mathcal{L}^{\text{Bog.}}\left( n\right) $.

The doubled up input-output map is of the form (\ref{G-io-t}), where now  $\tilde D= \tilde{S}$. The transfer function is explicitly
 \begin{equation}
\tilde \Xi_G(s) =  \left[  \begin{array}{c|c}
\tilde A  &  -\tilde C^\flat \tilde{S}
\\
\hline
\tilde C & \tilde{S}
\end{array}
\right] (s)= -\tilde C(sI-\tilde A)^{-1}\tilde C^\flat \tilde{S} + \tilde{S}.
\label{G-io-s-gen}
\end{equation}

The transfer function has the following properties:
\begin{enumerate}
\item  $\tilde{\Xi}_G\equiv \left[
\begin{tabular}{r|r}
$\tilde{A}$ & $-\tilde{C}^{\flat }$ \\ \hline
$\tilde{C}$ & $I$%
\end{tabular}
\right] \tilde{S}$;

\item  Whenever its value exists, we have $\tilde \Xi_G( i\omega )
\in \mathrm{Sp} ( \mathbb{C}^n) $, for $\omega \in \mathbb{R}$.
\end{enumerate}

Property 1 follows directly from (\ref{G-io-s-gen}) while Property
2 follows {\em mutatis mutandis} from the proof of \cite[Lemma
2]{GGY08} by the replacing $^{\dag}$ with $^{\flat}$ and
$(A,B,C,D)$ with $(\tilde A,\tilde B,\tilde C,\tilde D)$.

Physically, the meaning of (\ref{G-S-def-0}) is that to process an input signal, the Bogoliubov transformation $\tilde{S}$ is applied and the result is fed into the dynamical subsystem. 
As we have
argued in section \ref{sec:components-bog-comp}, Shale's theorem precludes   a unitary stochastic dynamical
model giving rise to a system  $\tilde{G}\in \mathcal{L}^{\text{Bog.}}\left(
n\right) $. Nevertheless, as we have also seen that there will exist a sequences $%
\tilde{G}_{L}\in \mathcal{L}^{\text{HP}}\left( n\right) $ such that
pointwise
\begin{equation*}
\lim_{L\rightarrow \infty }\Xi _{\tilde{G}_{L}}\left( s\right) =\Xi _{\tilde{%
G}}\left( s\right) ,
\end{equation*}
with $\tilde{G}\in \mathcal{L}^{\text{Bog.}}\left( n\right) $ but not in $%
\mathcal{L}^{\text{HP}}\left( n\right) $. One might envisage other modes of
convergence of transfer function, however, we shall restrict to pointwise
convergence for the purposes of this paper. It is interesting to note that
the class of Hudson-Parthasarathy models is not closed in the above sense of
convergence in the input-output sense, but may be extended to include
Bogoliubov transformations.

We remark that in many cases where a boson field is in a squeezed state (recall section \ref{sec:components-bog-aw}), this field may be regarded as the output of a static Bogoliubov component $\tilde S$ driven by vacuum inputs. This means, for example, that a dynamical component $(1, \tilde C, \tilde \Omega)$ driven by squeezed fields may be represented as a dynamical Bogoliubov component $(\tilde S, \tilde C, \tilde \Omega)$.

\subsection{Example: Cavity with Squeezed Input}
\label{sec:eg-2-cavity-sq}

Consider the cavity discussed in section \ref{sec:eg-2-cavity}, where now
we suppose that the cavity input is given by the output of a squeezer $G_\text{sq}$, described by 
the  Bogoliubov transformation
\begin{equation}
\tilde S_\text{sq} = \Delta( \cosh \lambda, \sinh \lambda )
=
\left[ \begin{array}{cc}
\cosh r & \sinh r
\\
\sinh r  & \cosh r
\end{array} \right].
\label{squeezer}
\end{equation}
That is,
$$
G_\text{sq} = \left(
 \Delta( \cosh \lambda, \sinh \lambda ),  0, 0
\right).
$$

The squeezed-input cavity $G_\text{cav, sq} = \tilde G_\text{cav} \triangleleft \tilde G_\text{sq}$ has transfer function
\begin{eqnarray}
\tilde \Xi_\text{cav, sq}(s) &=&  \tilde \Xi_\text{cav}(s) \tilde S_\text{sq}
\nonumber \\
&=& 
\left[ \begin{array}{cc}
\frac{s-\frac{\gamma}{2} +i \omega }{s+\frac{\gamma}{2} +i \omega} \cosh r & 
\frac{s-\frac{\gamma}{2} +i \omega }{s+\frac{\gamma}{2} +i \omega} \sinh r
\\
\frac{s-\frac{\gamma}{2} - i \omega  }{s+\frac{\gamma}{2} -i \omega} \sinh r
& \frac{s-\frac{\gamma}{2} - i \omega  }{s+\frac{\gamma}{2} -i \omega}  \cosh r
\end{array} \right].
\label{cav-sq-1}
\end{eqnarray}
This corresponds to the equations
\begin{eqnarray}
\left[  \begin{array}{c}
\dot a
\\
\dot a^\ast
\end{array} \right]
&=&
\left[  \begin{array}{cc}
-(\frac{\gamma}{2}+i \omega ) & 0
\\
0 & -(\frac{\gamma}{2}-i \omega ) 
\end{array} \right]
\left[  \begin{array}{c}
a
\\
 a^\ast
\end{array} \right] 
\nonumber \\ &&
+ \left[  \begin{array}{cc}
\cosh  r & \sinh  r
\\
\sinh  r  & \cosh  r 
\end{array} \right]
\left[  \begin{array}{c}
 b
\\
b^\ast
\end{array} \right]
\nonumber 
\\
\left[ \begin{array}{c}
 b_{out}
\\
b_{out}^\ast
\end{array} \right]
&=&
\left[ \begin{array}{cc}
\sqrt{\gamma}  & 0
\\
0 &\sqrt{\gamma} 
\end{array} \right]
\left[  \begin{array}{c}
a
\\
 a^\ast
\end{array} \right]
\nonumber
\\ &&
+\left[  \begin{array}{cc}
\cosh  r & \sinh  r
\\
\sinh  r  & \cosh  r
\end{array} \right]
\left[  \begin{array}{c}
b
\\
 b^\ast
\end{array} \right] .
\label{G-sc-eqns-1}
\end{eqnarray}

The physical parameters for the squeezed-input cavity are $\Omega_\text{cav,sq\,-}=\omega$, $\Omega_\text{cav,sq\,+}=0$, $C_\text{cav,sq\,-}=\sqrt{\gamma}$, $C_\text{cav,sq\,+}=0$, $\tilde S_\text{cav,sq\,-}= \tilde S_\text{sq} = \Delta(\cosh r, \sinh r)$, and so
$$
G_\text{cav, sq} = ( \Delta( \cosh \lambda, \sinh \lambda ), \Delta( \sqrt{\gamma}\, I, 0), -i \Delta(i \omega , 0)) .
$$
This system is a member of $\mathcal{L}^{\text{Bog.}}(1)$ but not of $\mathcal{L}^{\text{HP}}(1)$.

%?? transfer function

\section{Linear Quantum Feedback Networks}
\label{sec:lqfn}

\begin{figure}[h]
\begin{center}

\setlength{\unitlength}{1973sp}%
\begingroup\makeatletter\ifx\SetFigFont\undefined%
\gdef\SetFigFont#1#2#3#4#5{%
 \reset@font\fontsize{#1}{#2pt}%
 \fontfamily{#3}\fontseries{#4}\fontshape{#5}%
 \selectfont}%
\fi\endgroup%
\begin{picture}(7244,4150)(1479,-5689)
\put(4951,-4936){\makebox(0,0)[lb]{\smash{{\SetFigFont{6}{7.2}{\familydefault}{\mddefault}{\updefault}{\color[rgb]{0,0,0}$\Theta_\tau$}%
}}}}
\thicklines
{\color[rgb]{0,0,0}\put(7801,-3061){\vector(-1, 0){1500}}
\put(7801,-3061){\line( 0,-1){1800}}
\put(7801,-4861){\line(-1, 0){2100}}
}%
{\color[rgb]{0,0,0}\put(4501,-4861){\line(-1, 0){2100}}
\put(2401,-4861){\line( 0, 1){1800}}
\put(2401,-3061){\line( 1, 0){1500}}
}%
{\color[rgb]{0,0,0}\put(3001,-2161){\vector(-1, 0){1500}}
}%
{\color[rgb]{0,0,0}\put(7801,-2161){\vector(-1, 0){1500}}
}%
{\color[rgb]{0,0,0}\put(2851,-2161){\line( 1, 0){1050}}
}%
{\color[rgb]{0,0,0}\put(7651,-2161){\line( 1, 0){1050}}
}%
{\color[rgb]{0,0,0}\put(4801,-5161){\framebox(600,600){}}
}%
{\color[rgb]{0,0,0}\put(4351,-4861){\vector( 1, 0){450}}
}%
{\color[rgb]{0,0,0}\put(5776,-4861){\line(-1, 0){375}}
}%
\put(5851,-2236){\makebox(0,0)[lb]{\smash{{\SetFigFont{6}{7.2}{\familydefault}{\mddefault}{\updefault}{\color[rgb]{0,0,0}$b_1$}%
}}}}
\put(4126,-3136){\makebox(0,0)[lb]{\smash{{\SetFigFont{6}{7.2}{\familydefault}{\mddefault}{\updefault}{\color[rgb]{0,0,0}$b_{\mathrm{out},2}$}%
}}}}
\put(4126,-2236){\makebox(0,0)[lb]{\smash{{\SetFigFont{6}{7.2}{\familydefault}{\mddefault}{\updefault}{\color[rgb]{0,0,0}$b_{\mathrm{out},1}$}%
}}}}
\put(5851,-3136){\makebox(0,0)[lb]{\smash{{\SetFigFont{6}{7.2}{\familydefault}{\mddefault}{\updefault}{\color[rgb]{0,0,0}$b_2$}%
}}}}
\put(4951,-2686){\makebox(0,0)[lb]{\smash{{\SetFigFont{6}{7.2}{\familydefault}{\mddefault}{\updefault}{\color[rgb]{0,0,0}$G$}%
}}}}
\put(4576,-5611){\makebox(0,0)[lb]{\smash{{\SetFigFont{6}{7.2}{\familydefault}{\mddefault}{\updefault}{\color[rgb]{0,0,0}$\mathcal{F}_l(G,\Theta_\tau)$}%
}}}}
{\color[rgb]{0,0,0}\put(3901,-3661){\framebox(2400,2100){}}
}%
\end{picture}%

\caption{General form of a {\em  linear quantum feedback network (LQFN)}, with the time delay due to the spatial extent of the feedback connection represented by $\Theta_\tau$ (see text).}
\label{fig:lqfn}
\end{center}
\end{figure}
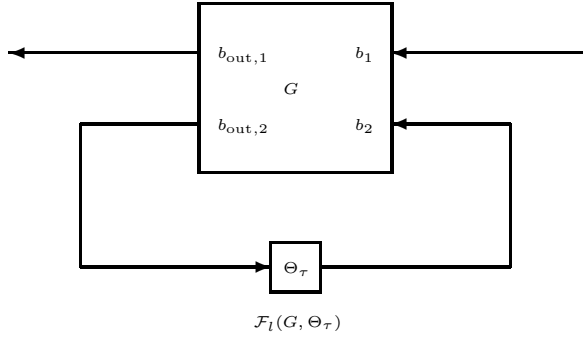

We are now in a position to described feedback networks constructed from Bogoliubov dynamical components as nodes, and boson fields as  links.  The general form of such a {\em linear quantum feedback network} (LQFN) is shown in Figure \ref{fig:lqfn}. The fundamental algebraic tool for describing such networks in subsequent sections is the linear fractional transformation (LFT).

\subsection{Linear Fractional Transformations}
\label{sec:lft}

Linear fractional transformations (LFTs) arise naturally when dealing with feedback networks, and a formal notation has been developed in classical linear systems theory, \cite{ZDG96}. Consider a classical transfer function $\Xi(s)$ partitioned as
$$
\Xi(s) = \left[  \begin{array}{cc}
\Xi_{11}(s) & \Xi_{12}(s)
\\
\Xi_{21}(s) & \Xi_{22}(s)
\end{array} \right]
$$
corresponding to a partition $u=(u_1,u_2)^\top$, $y=(y_1,y_2)^\top$ of the input and output signals. If the system is placed in a feedback arrangement defined by $u_2=K(s) y_2$, then the closed-loop system is described by the transfer function 
\begin{eqnarray*}
& \mathfrak{F}( \Xi(s), K(s) ) = 
\\
& \Xi_{11}(s)  + \Xi_{12}(s) K(s)  [I- \Xi_{22}(s) K(s) ]^{-1} \Xi_{21}(s).
\end{eqnarray*}
The arrangement is said to be well-posed whenever the inverse $ [I- \Xi_{22}(s) K(s) ]^{-1}$ exist. This transfer function is obtained by eliminating the in-loop variables.

In what follows we generalize this type of representation to our class of LQFNs (see also \cite{GJ08}, \cite{GGY08}).

\subsection{Fractional Linear Transformations}

In this section we provide some technical results needed for the network theory described in subsequent sections.

%\bigskip

%\noindent
%\textbf{Lemma}
\begin{lemma}
Let $\tilde{S} =\Delta (S_- ,S_+ ) \in \mathrm{Sp} ( \mathbb{C}^{n_{1}+n_{2}}) $ with block decomposition
$$ S_\mp = \left[
\begin{array}{cc}
S^\mp_{11} & S^\mp_{12}
\\
S^\mp_{21} & S^\mp_{22}
\end{array} \right],
$$
where $S_{jk}^\mp \in \mathbb{C}^{ n_j \times n_k }$. Setting $\hat S_{jk} =\Delta ( S^-_{jk}, S^+_{jk})\in \mathbb{C}^{ 2n_j \times 2n_k }$, we have that
\begin{equation}
\sum_{k=1,2}\hat{S}_{ki}^{\flat }\hat{S}_{kj}=\sum_{k=1,2}\hat{S}_{ik}\hat{S}%
_{jk}^{\flat }=\delta _{ij}.  \label{partitioned bog}
\end{equation}
\end{lemma}

%\textbf{proof:}
\begin{proof}
The relation $\tilde{S}^{\flat }\tilde{S}=I$ implies that $\Delta (S^{\dag }_-,-S^{\top
}_+)\Delta  ( S_{-},S_{+}) =\Delta  ( I,0) $, and so $%
S^{\dag }_- S_{-} - S^{\top }_ + S^{\#}_+=I$, $S^{\dag }_- S_{+}-S^{\top }_+S^{\#}_-=0$.
These may be written as $\sum_{k=1,2}(S_{ki}^{-\dag
}S_{kj}^{-}-S_{ki}^{+\top }S_{kj}^{+\#})=\delta _{ij}$, and $%
\sum_{k=1,2}(S_{ki}^{-\dag }S_{kj}^{+}-S_{ki}^{+\top }S_{kj}^{-\#})=0$.
Therefore
\begin{equation*}
\sum_{k=1,2}\hat{S}_{ki}^{\flat }\hat{S}_{kj} =\sum_{k=1,2}\Delta
(S_{ki}^{-\dag },-S_{ki}^{+\top })\Delta (S_{kj}^{-},S_{kj}^{+}) 
\end{equation*}
\begin{equation*}
=\sum_{k=1,2}\Delta (S_{ki}^{-\dag }S_{kj}^{-}-S_{ki}^{+\top
}S_{kj}^{+\#},S_{ki}^{-\dag }S_{kj}^{+}-S_{ki}^{+\top }S_{kj}^{-\#}) 
\end{equation*}
which equals $\Delta   ( \delta _{ij},0) =\delta _{ij}$. The second identity similarly follows from $\tilde{S} \tilde{S}^{\flat }=I$.
$\square $
\end{proof}

%\bigskip

%\noindent
%\textbf{Theorem}
\begin{theorem}
Let $\tilde{S}\in \mathrm{Sp} ( \mathbb{C}^{n_{1}+n_{2}}) $ and define
the fractional linear (M\"{o}bius) transformation $\Psi^{2 \to 1} _{\tilde{S}}:dom(\Psi^{2 \to 1}
_{\tilde{S}})\in \mathbb{C}^{n_{2}\times n_{2}}\mapsto \mathbb{C}^{n_{1}\times
n_{1}}$ by
\begin{equation}
\Psi^{2 \to 1} _{\tilde{S}} ( X) \triangleq \hat{S}_{11}+\hat{S}_{12}X (
I-
\hat{S}_{22}X )^{-1}\hat{S}_{21},
\label{Psi}
\end{equation}
with $X\in dom(\Psi^{2 \to 1} _{\tilde{S}})$ if and only if the inverse ($1-\hat{S}_{22}X)^{-1}
$ exists. Then $\Psi^{2 \to 1} _{\tilde{S}}$ maps $\mathrm{Sp} ( \mathbb{C}^{n_{2}})
\cap dom(\Psi^{2 \to 1} _{\tilde{S}})$ into $\mathrm{Sp} ( \mathbb{C}^{n_{1}}) $.
\end{theorem}
%\textbf{proof:}
\begin{proof}
We first note the Siegel-type identities
\begin{widetext}
\begin{eqnarray}
\Psi^{2 \to 1} _{\tilde{S}} ( X) ^{\flat }\Psi^{2 \to 1} _{\tilde{S}} ( Y)  &=&I-\hat{S}%
_{21}^{\flat }(I-X^{\flat }\hat{S}_{22}^{\flat })^{-1}(I-X^{\flat }Y)(I-%
\hat{S}_{22}Y)^{-1}\hat{S}_{21}, \nonumber \\ 
\Psi^{2 \to 1} _{\tilde{S}} ( X) \Psi^{2 \to 1} _{\tilde{S}} ( Y) ^{\flat } &=&I-\hat{S}%
_{12}(I-X\hat{S}_{22})^{-1}(I-XY^{\flat })(I-\hat{S}_{22}^{\flat }Y^{\flat
})^{-1}\hat{S}_{12}^\flat .
\end{eqnarray}
\end{widetext}

These are structurally the same as the standard Siegel identities based on
partitioning a unitary $\hat{S}$ but the involution $\flat $ replacing the
usual Hermitian involution \dag , see Theorem 21.16 and Corollary 21.17 of reference \cite{Young88}. The
identities rely on the unitary analogue of the identities $ ( \ref
{partitioned bog}) $ and so follow \emph{mutatis mutandis}. Evidently,
if $X\in \mathrm{Sp} ( \mathbb{C}^{n_{2}}) \cap dom(\Psi^{2 \to 1} _{\tilde{S}})$
then $\Psi^{2 \to 1} _{\tilde{S}} ( X) ^{\flat }\Psi^{2 \to 1} _{\tilde{S}} ( X) =\Psi^{2 \to 1}
_{\tilde{S}} ( X) \Psi^{2 \to 1} _{\tilde{S}} ( X) ^{\flat }=I$. 
$\square$
\end{proof}
\bigskip

\noindent
\begin{corollary}
If $\tilde{K} ( i\omega ) $ is an $\mathrm{Sp} ( 
\mathbb{C}^{n_{2}}) $-valued transfer matrix function taking values in $dom(\Psi^{2 \to 1} _{\tilde{S}})$ for all $\omega$ real, then the
fractional linear transformation $\Psi^{2 \to 1} _{\tilde{S}} ( \tilde{K} ( i\omega )
) $ will be $\mathrm{Sp} ( \mathbb{C}^{n_1}) $-valued function of $\omega$. 
%\bigskip

In particular, if 
$I\in dom(\Psi^{2 \to 1} _{\tilde{S}})$ then 
\begin{equation}
\Psi^{2 \to 1} _{\tilde{S}}(I)=
\hat{S}_{11}+\hat{S}_{12}(1-\hat{S}_{22})^{-1}\hat{S}_{21} \in \mathrm{Sp} ( \mathbb{C}^{n_1})
\label{W_0}.
\end{equation}
\end{corollary}

\subsection{Finite Time-Delay LQFNs}
\label{sec:gen-nw-defs}

A general LQFN is a network of linear quantum components $G_v \in \mathcal{L}^{\text{Bog.}}(n)$, labeled by the vertices $v$ of the network, with quantum fields traveling along the edges. The edges are directed so that we distinguish inputs and outputs, and the multiplicity of input fields equals the multiplicity of outputs for each component.

In a physical LQFN we will have time delays associated with each internal edge due to the finite time taken by light to travel from an output port to an input port. In fact, we may lump the individual components as one single global component $\tilde{G}$ with all external inputs going into a collective input port 1 and coming out from a collective output port 1, as in Figure \ref{fig:lqfn}. Likewise, all the internal fields can be viewed as traveling from the collective output port 2 to the collective input port 2. The effect of the (multichannel) time-delay can be described by the operator $\Theta_\tau$ defined by 
\begin{equation*}
\Theta_\tau (f_1(t), \ldots, f_n(t))^\top = (f_1(t-\tau_1), \ldots, f_n(t-\tau_n))^\top,
\end{equation*}
where $\tau_1 > 0, \ldots, \tau_n > 0$ are the time delays of each channel. Here $f_k(t)$ denotes  the quantum stochastic process propagating along channel $k$ in doubled-up form. For instance, $f_k(t)$ could be $\breve y_k(t)$, the doubled-up output quantum output processes  propagating along channel $k$. In a slight abuse of notation, we also occasionally overload the notation $\Theta_\tau$ %(as in Appendix A) 
to denote the delayed version of a quantum process that is not in doubled-up form, such as when $f_k(t)$ is taken to be $y_k(t)$ for all $k$. Note that $ [\Theta_\tau (i\omega )]_{jk} =e^{i\omega \tau_j} \delta_{jk}$.
Extending the standard notation recalled in section \ref{sec:lft}), we denote this by $$\tilde{N}_\tau =\mathfrak{F} (\tilde{G} ,\Theta_{\tau}).$$

A Hamiltonian for a LQFN with squeezing components could be constructed approximately by replacing Bogoliubov components $\tilde{S}$ with dynamical components $G_{\tilde{S}}^\epsilon$. This would then fit into the QFN framework of \cite{GJ08a}.

\subsection{Parameters for Network Model}
\label{sec:gen-nw-model}

We now suppose that the LQFN of Figure \ref{fig:lqfn} is described by field channels $b_1, b_{\mathrm{out},1}$ and $b_2, b_{\mathrm{out},2}$ having lengths $n_1$ and $n_2$, respectively, so that the total number is $n_1+n_2=n$. The system is parameterized by $G=(\tilde{S},L,H)$, with $\tilde{S}=\Delta(S_-, S_+)$ and we partition the matrices as
\begin{eqnarray*}
C_\mp = \left[ \begin{array}{c}
C^\mp_1
\\
C^\mp_2
\end{array} \right], 
\
S_\mp= \left[
\begin{array}{cc}
S^\mp_{11} & S^\mp_{12}
\\
S^\mp_{21} & S^\mp_{22}
\end{array} \right] .
\end{eqnarray*}
The field-field component of the input-output relations can be now written as
$$ \breve{b}_{\mathrm{out},i} = \sum_{j=1,2} \hat{G}_{ij} (s) \breve{b}_j ,$$
with transfer matrix function
\begin{eqnarray}
\hat \Xi_G(s) =  \left[  \begin{array}{c|c} \tilde A  &  - [ \tilde
C_1^\flat, \tilde C_2^\flat ] \hat{S}
\\
\hline \left[ \begin{array}{c} \tilde C_1
\\
\tilde C_2
\end{array} \right]
& \hat{S}
\end{array}
\right] (s) \nonumber
\\
= -\left[ \begin{array}{c} \tilde C_1
\\
\tilde C_2
\end{array} \right]
(sI-\tilde A)^{-1}
[\begin{array}{c} 
\tilde C_1^\flat, \tilde C_2^\flat  \\
\end{array}]
\hat{S} +\hat{S}
\label{G-io-lqfn}
\end{eqnarray}
where
$$
\hat{S}_{jk} =
\Delta(S^-_{jk}, S^+_{jk}) , \ \ \tilde C_j = \Delta(C^-_j,  C^+_j).
$$

The network $N_\tau$ is given by the  linear fractional transformation
\begin{equation}
\tilde{N}_\tau =\mathfrak{F} (\tilde{G} ,\Theta_{\tau}) =  \left[  \begin{array}{c|c}
\tilde A_\tau  &  -\tilde C_\tau^\flat \tilde{S}_\tau
\\
\hline \tilde C_\tau & \tilde{S}_\tau
\end{array}
\right]
\label{N-lft}
\end{equation}
where
\begin{eqnarray}
\tilde{S}_\tau &=& \hat{S}_{11} + \hat{S}_{12} \Theta_\tau (I - \hat{S}_{22} \Theta_\tau )^{-1} \hat{S}_{21}
\label{N-lft-W}
 \\
\tilde C_\tau &=&  \tilde C_1 + \hat{S}_{12}\Theta_\tau  (I - \hat{S}_{22}\Theta_\tau )^{-1}  \tilde C_2
\label{N-lft-C}
 \\
\tilde A_\tau &=&  \tilde A - \sum_{j=1,2} \tilde C_j^\flat \hat{S}_{j2} \Theta_\tau (I - \hat{S}_{22}\Theta_\tau)^{-1} \tilde C_2 .
 \label{N-lft-A}
\end{eqnarray}
Due to the non-zero delay, the network model $N_\tau$ is non-Markovian.

\subsection{Zero Delay Limit Models}
\label{sec:gen-nw-zero}

Of particular interest are the simpler models that arise in the
zero-delay limit $\Theta_\tau \to I$ ($\tau \to 0$). Assume that
$I-\hat{S}_{11}$ is invertible. From above we have
\begin{equation}
\tilde{N}_0 = \mathfrak{F}(\tilde{G},I) =  \left[  \begin{array}{c|c}
\tilde A_0  &  -\tilde C_0^\flat \tilde{S}_0
\\
\hline \tilde C_0 & \tilde{S}_0
\end{array}
\right]
\label{N-lft-I}
\end{equation}
where
\begin{eqnarray}
\tilde{S}_0 &=& \hat{S}_{11} + \hat{S}_{12} (I - \hat{S}_{22} )^{-1} \hat{S}_{21 }
\label{N-lft-S-I}
 \\
\tilde C_0 &=&  \tilde C_1 + \hat{S}_{12}  (I - \hat{S}_{22} )^{-1}  \tilde C_2
\label{N-lft-C-I}
 \\
\tilde A_0 &=&  \tilde A - \sum_{j=1,2} \tilde C_j^\flat \hat{S}_{j2}   (I - \hat{S}_{22} )^{-1} \tilde C_2 .
 \label{N-lft-A-I}
\end{eqnarray}
We note that
\begin{equation}
\tilde A_0 = -\frac{1}{2} \tilde C_0^\flat \tilde C_0 - i \tilde \Omega_0,
\label{N-lft-A0}
\end{equation}
where
\begin{equation}
\tilde \Omega_0 = \tilde \Omega +  \mathrm{Im}_\flat \sum_{j=1,2} \tilde
C_j^\flat \hat{S}_{j2} (I-\hat{S}_{22})^{-1} \tilde C_2 .
\label{N-lft-Omega0}
\end{equation}
Here, $\mathrm{Im}_\flat X$ means $\frac{1}{2i}( X-X^\flat)$. The matrix $\tilde{S}_0$ defined by (\ref{N-lft-S-I}) is a Bogoliubov matrix, as it corresponds to the matrix in (\ref{W_0}). 
Therefore the zero-delay limit $N_0$ is a Markovian system belongs to $\mathcal{L}^{\text{Bog.}}(n)$ with parameters
\begin{eqnarray}
N_0 &=& \left( { \hat{S}_{11} + \hat{S}_{12} (I - \hat{S}_{22} )^{-1} \hat{S}_{21 } ,} \right.
\nonumber \\
&& \tilde C_1 + \hat{S}_{12}  (I - \hat{S}_{22} )^{-1}  \tilde C_2 ,
\nonumber \\
&& \left. {\tilde \Omega +  \mathrm{Im}_\flat \sum\nolimits_{j=1,2} \tilde
C_j^\flat \hat{S}_{j2} (I-\hat{S}_{22})^{-1} \tilde C_2 } \right) .
\label{N0-params}
\end{eqnarray}
Thus $\mathcal{L}^{\text{Bog.}}(n)$ is closed with respect to this zero-delay limit network construction.

Other types of  limits are also considered in applications (see, e.g. \cite[sec. 2.3]{BR04}).
Suppose that the system $G^\epsilon$ and the delay $\tau^\epsilon$ depend on a small parameter
$\epsilon > 0$, defining a physical regime of operation. Then one may obtain a limit model
$\lim_{\epsilon\to 0} \mathfrak{F}(G^\epsilon, \Theta_{\tau^\epsilon})$. An example of this is
considered in section \ref{sec:eg-2-dyn} below.

\subsection{Series Product}
\label{sec:gen-nw-series}

The series product $G_2 \triangleleft G_1$ of  two systems $\tilde
G_1=(\tilde{S}_1, \tilde C_1, \tilde \Omega_1)$ and $\tilde G_1=(\tilde{S}_2, \tilde
C_2, \tilde \Omega_2)$ follows from the zero-delay limit
(\ref{N-lft-I}). For the series product, we interchange the index $1$ and $2$ (this simply 
means interchanging the role of $G_1$ and $G_2$ in the LQFN) and 
then set $\hat{S}_{12}=\tilde{S}_1$, $\hat{S}_{21}=
\tilde{S}_2$, $\hat{S}_{22}=0$ and $\hat{S}_{11}=0$ (note that without the interchange
we would be computing $ G_1 \triangleleft G_2$ instead $ G_2 \triangleleft G_1$). 

Substituting into (\ref{N-lft-S-I},\ref{N-lft-C-I},\ref{N-lft-A-I}) we find
\begin{equation}
\tilde \Xi _{\text{series}}=\left[ 
\begin{tabular}{r|r}
$\tilde A - \tilde C_{2}^{\flat } \tilde S_{2} \tilde C_{1}$ & $-\left( \tilde C_{2}^{\flat
}\tilde S_{2}+\tilde C_{1}^{\flat }\right) \tilde S_{1}$ \\ \hline
$\tilde C_{2}+\tilde S_{2}\tilde C_{1}$ & $\tilde S_{2} \tilde S_{1}$%
\end{tabular}
\right] .
\label{seriesTF}
\end{equation}

The matrices  for $G_2
\triangleleft G_1 =(\tilde{S}_{\mathrm{series}}, \tilde
C_{\mathrm{series}}, \tilde \Omega_{\mathrm{series}})$ are given by
\begin{eqnarray*}
\tilde{S}_{\mathrm{series}} &=&\tilde{S}_{2}\tilde{S}_{1},  \notag \\
\tilde{C}_{\mathrm{series}} &=&\tilde{C}_{2}+\tilde{S}_{2}\tilde{C}_{1} =
\Delta{( C_{\mathrm{series}-}, C _{\mathrm{series}+})
},
\notag \\
\tilde{\Omega}_{\mathrm{series}} &=&\tilde{\Omega}_1 + \tilde\Omega_2 +\mathrm{Im}_{\flat }\tilde{C}%
_{2}^{\flat }\tilde{S}_{2}\tilde{C}_{1}
\notag
\\ & \equiv& \Delta{( \Omega _{\mathrm{%
series}-}, \Omega_{\mathrm{series}+}) },
\end{eqnarray*}
where
\begin{eqnarray*}
C_{\mathrm{series}-} &=&C_{2-}+S_{2-}C_{1-}+ S_{2+} C _{1+}^{\#}, \\
C_{\mathrm{series}+} &=& C_{2+}+S_{2-} C_{1+}+ S_{2+}C_{1-}^{\#}.
\end{eqnarray*}
From
\begin{eqnarray*}
\tilde{C}_{2}^{\flat }\tilde{S}_{2}\tilde{C}_{1}&=&\Delta (
C_{2-}^{\dag },- C_{2+}^{\top }) \Delta{( S_{2-},
S_{2+}) }\Delta{( C_{1-},C_{1+})
}\\
&=&\Delta{( X_-,X_+ ) }
\end{eqnarray*}
with
\begin{equation*}
X_\mp = ( C_{2-}^{\dag }S_{2-}-C_{2+}^{\top
}S_{2+}^{\#})C_{1\mp}+ (C_{2-}^{\dag } S_{2+} - C_{2+}^{\top}S_{2-}^{\#}) C_{1\pm}^{\#}, 
\end{equation*}
we see that
\begin{eqnarray*}
\Omega _{\mathrm{series}-}&=&\Omega_{1-}+\Omega
_{2-} +\frac{1}{2i}( X_- -X_-^\dag ),\\
\Omega_{\mathrm{series}+}&=& \Omega_{1+}+\Omega_{2+}+\frac{1}{2i}
( X_+ +X_+^{\top } ).
\end{eqnarray*}

Succinctly, the series product in $\mathcal{L}^{\text{Bog.}}(n)$ is given by
\begin{equation}
G_2 \triangleleft G_1 = (\tilde S_2 \tilde S_1, \tilde C_2+ \tilde S_2 \tilde C_1, \tilde \Omega_1+\tilde \Omega_2 + \mathrm{Im}_\flat \{  \tilde C_2^\flat \tilde S_2 \tilde C_1 \} );
\label{HP-series-plus}
\end{equation}
cf. (\ref{HP-series-def}).

Clearly, $\mathcal{L}^{\text{Bog.}}(n)$ is a group with respect to the series product, and the classes of components  $\mathcal{L}^{\text{HP}}(n)$,  $\mathrm{Sp} ( \mathbb{C}^n) $, and $U(n)$ are subgroups. If $G=(\tilde S, \tilde C, \tilde\Omega)$ the inverse is given by 
\begin{equation}
G^{-1}=(\tilde S^\flat, -\tilde S^\flat \tilde C, -\tilde\Omega),
\label{series-inverse}
\end{equation} 
with transfer function given by
\begin{equation}
\tilde{\Xi}_{G^{-1}}\left( s\right) \equiv \tilde{\Xi}_{G}\left( s^{\ast }\right) ^{\flat }.
\label{G-1:tf}
\end{equation}

The series product is not limited simply to feedforward, and the above formula apply to the case where the two systems have one or more modes in common. The series product is therefore highly non-trivial \cite{GJ08,GJ08a,GGY08}. 

\subsection{Series Product and Cascaded Transfer Functions}
\label{sec:gen-nw-series-tf}

In classical linear systems theory, the transfer function of a cascade of two separate systems is obtained by multiplying the transfer functions (see, e.g., \cite{ZDG96}).

We need to emphasize here that two systems will be distinct if they consist
of different oscillator modes. Specifically let there be $m_{1}$ modes in
the first system and $m_{2}$ in the second, and set $a=\left[ 
\begin{array}{c}
a_{1} \\ 
a_{2}
\end{array}
\right] \in \mathcal{S}\left( m_{1}+m_{2}\right) $, then we wish to consider 
$G_{1}=\left( \tilde{S}_{1},\tilde{C}_{1},\tilde{\Omega}_{1}\right) $ and $%
G_{2}=\left( \tilde{S}_{2},\tilde{C}_{2},\tilde{\Omega}_{2}\right) $ with
\begin{eqnarray*}
\tilde{C}_{1} &=&\Delta \left( \left[ C_{1-},0\right] ,\left[ C_{1+},0\right]
\right) , \\
-i\tilde{\Omega}_{1} &=& -\Delta \left( \left[ 
\begin{array}{cc}
i\Omega _{1-} & 0 \\ 
0 & 0
\end{array}
\right] ,\left[ 
\begin{array}{cc}
i\Omega _{1+} & 0 \\ 
0 & 0
\end{array}
\right] \right) , \\
\tilde{C}_{2} &=&\Delta \left( \left[ 0,C_{2-}\right] ,\left[ 0,C_{2+}\right]
\right) , \\
-i\tilde{\Omega}_{2} &=& - \Delta \left( \left[ 
\begin{array}{cc}
0 & 0 \\ 
0 & i\Omega _{2-}
\end{array}
\right] ,\left[ 
\begin{array}{cc}
0 & 0 \\ 
0 & i\Omega _{2+}
\end{array}
\right] \right) ,
\end{eqnarray*}
with respect to the decomposition $\mathbb{C}^{m}=\mathbb{C}^{m_{1}}\oplus 
\mathbb{C}^{m_{2}}$.  This is simply a statement that the dynamics  of $G_1$ does not depend on the internal variables of $G_2$, and vice-versa. Putting this particular form into (\ref{seriesTF}), we then obtain
\begin{eqnarray}
\tilde{\Xi}_{G_{2}\triangleleft G_{1}} (s) &=& 
\left[ 
\begin{tabular}{c|c}
$\left[
\begin{array}{cc}
\bar A_{1} & 0 \\ 
-\bar C_{2}^{\flat }\tilde S_{2}\bar C_{1} & \bar A_{2}
\end{array}
\right] $ & $\left[
\begin{array}{c}
-\bar C_{1}^{\flat }\tilde S_{1} \\ 
-\bar C_{2}^{\flat }\tilde S_{2}\tilde S_{1}
\end{array}
\right] $ \\ \hline
$\left[ \tilde S_2 \bar C_{1},\bar C_{2}\right] $ & $\tilde S_{2}\tilde S_{1}$%
\end{tabular}
\right] (s) \notag \\
&=& \tilde \Xi_{G_2}
(s) \ \tilde \Xi_{G_1} (s),
\label{cascade}
\end{eqnarray}
where
$\bar C_j =\Delta ( C_{j,-} ,C_{j,+} )$ and $\bar A_j = -\frac{1}{2} \bar C_j^\flat \bar C_j - i \Delta ( \Omega_{j,-} , \Omega_{j,+} )$.
Here we use the unitary transformation $\breve a \mapsto \left[ 
\begin{array}{c}
\breve a_{1} \\ 
\breve a_{2}
\end{array}
\right]$ to present the transfer function in a more convenient form. The algebra is then similar to Section IV A in \cite{GGY08}.

If the systems are not separate, then we do not expect such a factorization of the transfer function to hold.
The series product \cite{GJ08a}, \cite{GJ08},  \cite{GGY08} is defined quite generally in terms of physical parameters (section \ref{sec:components-dyn-series}) which may, for example, depend on the same oscillator mode variables).  In the general case, the transfer function can be computed using the general formulas (\ref{N-lft-I})-(\ref{N-lft-A-I}).

Let us remark that the series product inverse $G^{-1}$ given by (\ref{series-inverse}) may be realized in terms of a physical system that
is  \emph{not} separate from the original system. Physically, if we pass input fields through a system with parameters $G$, then $G^{-1}$ gives the parameters required to undo the effect by passing the output back through the \emph{same} system for a second pass.

\subsection{Inverse Transfer Functions}

The input-output relation $\breve{b}_{\text{out}}=\tilde{\Xi}\breve{b}_{\text{%
in}}+\tilde{\xi}\breve{a}\left( 0\right) $ may be inverted to yield 
\begin{equation*}
\breve{b}_{\text{in}}=\tilde{\Xi}^{-1}\breve{b}_{\text{out}}-\tilde{\Xi}^{-1}%
\tilde{\xi}\breve{a}\left( 0\right) .
\end{equation*}
In particular, we can give the following useful description of $\tilde{\Xi}%
^{-1}$. The linear equations (\ref{lin-dyn-d2}) in the time domain may be rearranged algebraically to give 
\begin{eqnarray*}
\frac{d}{dt}\breve{a} &=&(\tilde{A}-\tilde{B}\tilde{D}^{-1}\tilde{C})%
\breve{a}+\tilde{B}\tilde{D}^{-1}\breve{b}_{\text{out}} ,\\
\breve{b}_{\text{in}} &=&-\tilde{D}^{-1}\tilde{C}\breve{a}+\tilde{D}^{-1}%
\breve{b}_{\text{out}},
\end{eqnarray*}
with $\tilde{D}$  invertible, and in the transform domain, we deduce
that 
\begin{equation*}
\left[ 
\begin{tabular}{l|l}
$\tilde{A}$ & $\tilde{B}$ \\ \hline
$\tilde{C}$ & $\tilde{D}$%
\end{tabular}
\right] ^{-1}=\left[ 
\begin{tabular}{l|l}
$\tilde{A}-\tilde{B}\tilde{D}^{-1}\tilde{C}$ & $\tilde{B}\tilde{D}^{-1}$ \\ 
\hline
$-\tilde{D}^{-1}\tilde{C}$ & $\tilde{D}^{-1}$%
\end{tabular}
\right] .
\end{equation*}

For the model with parameters $G=(\tilde{S},\tilde{C},\tilde{\Omega})$, we
find 
\begin{eqnarray*}
\tilde{\Xi}_{G}^{-1}(s) &=& \left[ 
\begin{tabular}{l|l}
$\tilde{A}$ & $-\tilde{C}^{\flat }\tilde{S}$ \\ \hline
$\tilde{C}$ & $\tilde{S}$%
\end{tabular}
\right] ^{-1}\equiv \left[ 
\begin{tabular}{l|l}
$-\tilde{A}^{\flat }$ & $-\tilde{C}^{\flat }$ \\ \hline
$-\tilde{S}^{\flat }\tilde{C}$ & $\tilde{S}^{\flat }$%
\end{tabular}
\right] \\
&=&  \tilde S^\flat \tilde C(sI + \tilde A^\flat )^{-1}\tilde C^\flat  + \tilde{S}^\flat , 
\end{eqnarray*}
or 
\begin{equation}
\tilde{\Xi}_{G}\left( s\right) ^{-1}\equiv \tilde{\Xi}_{G}\left( -s^{\ast
}\right) ^{\flat }.
\end{equation}
We note that $
\tilde{\Xi}_{G^{-1}}\left( s\right) =\tilde{\Xi}_{G}\left( -s\right)
^{-1}\equiv \tilde{\Xi}_{G}\left( s^{\ast }\right) ^{\flat }$ (recall (\ref{G-1:tf})).

\subsubsection{Example: separate cavity inverse}
\label{sec:sep-cav-inv}

As a concrete example,  consider the single mode cavity $G=G_{cav}$ considered in \ref{sec:eg-2-cavity}. The transfer function (\ref{cavity-2}) and related functions are given by
\begin{eqnarray*}
\Xi_{G,-}(s)&=& \frac{s- \frac{\gamma}{2}+i\omega}{s+ \frac{\gamma}{2}+i\omega},\\
\Xi_{G^{-1},-}(s) &=& \frac{s- \frac{\gamma}{2}-i\omega}{s+ \frac{\gamma}{2}-i\omega},\\
\Xi_{G,-}^{-1}(s) &=& \frac{s+ \frac{\gamma}{2}+i\omega}{s- \frac{\gamma}{2}+i\omega}.
\end{eqnarray*}
That is, $G^{-1}$ is obtained from $G$ by keeping $C_- = \sqrt \gamma$, $C_+ =0$ and replacing $\Omega_- =\omega$ by $-\omega$.  In what follows we obtain a physical realization $\hat G$ of 
the transfer function $\Xi_{G,-}^{-1}(s)$ that is a system distinct from $G$, so that
\begin{equation}
\tilde \Xi_{\hat G}(s)=
\tilde \Xi_{G}^{-1}(s) .
\label{M-tf}
\end{equation}
The system $\hat G$ is obtained by  setting $\Omega_- = -\omega, \Omega_+ =0$ and swapping $C_- =0 $ and $C_+ = \sqrt{\gamma}$.

In this example we can take the two modes to be 
$\left[ 
\begin{array}{c}
a_{1} \\ 
a_{2}
\end{array}
\right] \in \mathcal{S}\left( 2\right) $ and write $G$ as $\left( \tilde{S}%
_{1}=I,\tilde{C}_{1},\tilde{\Omega}_{1}\right) $ where
\begin{eqnarray*}
\tilde{C}_{1}&=& \Delta \left( \left[ \sqrt{\gamma },0\right] ,\left[ 0,0\right]
\right) ,\\
-i\tilde{\Omega}_{1}& =& \Delta \left( \left[ 
\begin{array}{cc}
-i\omega  & 0 \\ 
0 & 0
\end{array}
\right] ,\left[ 
\begin{array}{cc}
0 & 0 \\ 
0 & 0
\end{array}
\right] \right) ,
\end{eqnarray*}
and $\hat G$ as $\left( \tilde{S}_{2}=I,\tilde{C}_{2},-\tilde{\Omega}_{2}\right) $
where
\begin{eqnarray*}
\tilde{C}_{2} &=& \Delta \left( \left[ 0,0\right] ,\left[ 0,\sqrt{\gamma }\right]
\right) ,\\
-i\tilde{\Omega}_{2} &=& -\Delta \left( \left[ 
\begin{array}{cc}
-i\omega  & 0 \\ 
0 & 0
\end{array}
\right] ,\left[ 
\begin{array}{cc}
0  & 0 \\ 
0 & 0
\end{array}
\right] \right) .
\end{eqnarray*}
That is, in terms of scattering matrices, coupling operators and Hamiltonians (\ref{G-params-0}) we have
$$
G = (I, \sqrt{\gamma}\, a_1, \omega a_1^\ast a_1), \ \mathrm{and} \  \hat G = (I, \sqrt{\gamma}\, a_1^\ast, -\omega a_2^\ast a_2).
$$

The series product $\hat G \triangleleft G$  is  given by $\left( I,%
\tilde{C}=\tilde{C}_{1}+\tilde{C}_{2},\tilde{\Omega}=\tilde{\Omega}_{1}+%
\tilde{\Omega}_{2}  + \mathrm{Im}_\flat[ \tilde C_2^\flat \tilde C_1] \right) $ (recall (\ref{HP-series-plus})) and this corresponds to a system with a
nontrivial dynamics, since (by (\ref{HP-series-def})
$$
\hat G \triangleleft G = (I, \sqrt{\gamma}\,(a_1+a_2^\ast), \omega(a_1^\ast a_1 - a_2^\ast a_2) + \mathrm{Im}[ a_2 a_1 ]).
$$
Nevertheless,  some calculation shows that
\begin{equation*}
\tilde{\Xi}_{\hat G \vartriangleleft G}\left( s\right) =I,
\end{equation*}
as required by (\ref{cascade}) and (\ref{M-tf}). This continues to hold if there is a term $\Omega_+ = \omega_+$ added to the cavity Hamiltonian. 

\subsubsection{General separate system inverses}

We now show how to obtain a physical realization for $\Xi^{-1}_G(s)$ in the case where $\tilde \Omega =0$.
We have
\begin{eqnarray*}
{\tilde \Xi}_{G}^{-1}(s) &=& \tilde S^\flat + \tilde S^\flat \tilde C (sI -\frac{1}{2} \tilde C^\flat \tilde C )^{-1} \tilde C^\flat\\
&=& [I+ \tilde S^\flat (s - \frac{1}{2}\tilde C \tilde C^\flat )^{-1} \tilde C \tilde C^\flat ]\tilde S^\flat .
\end{eqnarray*}
Now $\tilde C \tilde C^\flat$ is not definite and in fact we have
\begin{eqnarray*}
\tilde C \tilde C^\flat &=& \Delta (C_- C_-^\dag -C_+ C_+^\dag , -C_- C_+^\top + C_+ C_-^\top )\\
&=& - \Delta (C_+ , C_- ) \Delta ( C_+^\dag , - C_-^\top ) \\
&\equiv & - \tilde K \tilde K^\flat ,
\end{eqnarray*}
where
\begin{equation}
\tilde K = \Delta (C_+ , C_- ).
\end{equation}
We therefore obtain
\begin{eqnarray*}
\tilde{\Xi}_{G}^{-1}(s) 
&=& [I- \tilde S^\flat (s + \frac{1}{2}\tilde K \tilde K^\flat )^{-1} \tilde K \tilde K^\flat ]\tilde S^\flat \\
&=& [I- \tilde S^\flat \tilde K (s + \frac{1}{2}\tilde K^\flat \tilde K  )^{-1} \tilde K^\flat ]\tilde S^\flat \\
&\equiv & \tilde \Xi_{\hat G} (s),
\end{eqnarray*}
where
\begin{equation}
\hat G = ( \tilde S^\flat ,  \tilde S^\flat \tilde K, 0).
\end{equation}
We note that this choice of $\hat G$ is not unique in producting a transfer function inverse inverse to ${\tilde \Xi}_{G}$. 
Finding an inverse when $\tilde \Omega \neq 0$ is more involved.

\subsubsection{Comments}

We have the mapping $G \mapsto \tilde \Xi_G$ from the group of $\mathcal{L}^{\text{Bog.}}(n)$ of systems parameters with series product to the group of matrix transfer functions. This mapping is not however a group homomorphism. Indeed, we typically have 
\begin{equation*}
\tilde \Xi_{G_2 \triangleleft G_1} (s) \neq \tilde \Xi_{G_2}(s) \tilde \Xi_{G_1}(s) ,
\end{equation*}
though equality - the cascade formula (\ref{cascade}) - holds when the systems are separate assemblies of oscillators.

We should also caution that, as we have seen in the example in \ref{sec:sep-cav-inv}, there are solutions for $G$ other than the trivial $G=\left( I,0,0\right) $ to the equation  $\tilde{\Xi}_{G}\left( s\right) =I$.

\section{Network Examples}
\label{sec:eg-2}

\subsection{In-loop Squeezing and Cavity as a Feedback Network}
\label{sec:eg-2-lqfn}

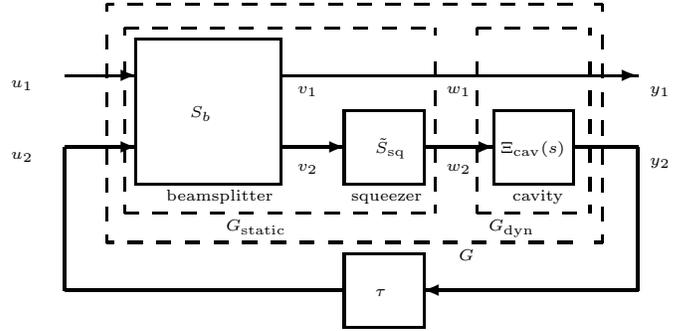
\begin{figure}[h]
\begin{center}

\setlength{\unitlength}{1973sp}%
\begingroup\makeatletter\ifx\SetFigFont\undefined%
\gdef\SetFigFont#1#2#3#4#5{%
 \reset@font\fontsize{#1}{#2pt}%
 \fontfamily{#3}\fontseries{#4}\fontshape{#5}%
 \selectfont}%
\fi\endgroup%
\begin{picture}(8055,4094)(511,-4583)
\put(6151,-3736){\makebox(0,0)[lb]{\smash{{\SetFigFont{6}{7.2}{\familydefault}{\mddefault}{\updefault}{\color[rgb]{0,0,0}$G$}%
}}}}
\thicklines
{\color[rgb]{0,0,0}\put(4726,-2761){\framebox(975,900){}}
}%
{\color[rgb]{0,0,0}\put(2101,-2761){\framebox(1800,1800){}}
}%
{\color[rgb]{0,0,0}\put(3901,-2311){\vector( 1, 0){825}}
}%
{\color[rgb]{0,0,0}\put(5701,-2311){\vector( 1, 0){900}}
}%
{\color[rgb]{0,0,0}\put(4726,-4561){\framebox(975,900){}}
}%
{\color[rgb]{0,0,0}\put(7576,-2311){\line( 1, 0){825}}
\put(8401,-2311){\line( 0,-1){1800}}
\put(8401,-4111){\vector(-1, 0){2700}}
}%
{\color[rgb]{0,0,0}\put(1201,-1411){\vector( 1, 0){900}}
}%
{\color[rgb]{0,0,0}\put(3901,-1411){\vector( 1, 0){4500}}
}%
{\color[rgb]{0,0,0}\put(6376,-3136){\dashbox{180}(1425,2325){}}
}%
{\color[rgb]{0,0,0}\put(4726,-4111){\line(-1, 0){3525}}
\put(1201,-4111){\line( 0, 1){1800}}
\put(1201,-2311){\vector( 1, 0){900}}
}%
{\color[rgb]{0,0,0}\put(1951,-3136){\dashbox{180}(3900,2325){}}
}%
{\color[rgb]{0,0,0}\put(1726,-3511){\dashbox{180}(6225,3000){}}
}%
\put(5101,-2386){\makebox(0,0)[lb]{\smash{{\SetFigFont{6}{7.2}{\familydefault}{\mddefault}{\updefault}{\color[rgb]{0,0,0}$\tilde S_\text{sq}$}%
}}}}
\put(8551,-2536){\makebox(0,0)[lb]{\smash{{\SetFigFont{6}{7.2}{\familydefault}{\mddefault}{\updefault}{\color[rgb]{0,0,0}$y_2$}%
}}}}
\put(8551,-1636){\makebox(0,0)[lb]{\smash{{\SetFigFont{6}{7.2}{\familydefault}{\mddefault}{\updefault}{\color[rgb]{0,0,0}$y_1$}%
}}}}
\put(526,-2461){\makebox(0,0)[lb]{\smash{{\SetFigFont{6}{7.2}{\familydefault}{\mddefault}{\updefault}{\color[rgb]{0,0,0}$u_2$}%
}}}}
\put(2776,-1936){\makebox(0,0)[lb]{\smash{{\SetFigFont{6}{7.2}{\familydefault}{\mddefault}{\updefault}{\color[rgb]{0,0,0}$S_b$}%
}}}}
\put(4126,-2611){\makebox(0,0)[lb]{\smash{{\SetFigFont{6}{7.2}{\familydefault}{\mddefault}{\updefault}{\color[rgb]{0,0,0}$v_2$}%
}}}}
\put(6001,-2611){\makebox(0,0)[lb]{\smash{{\SetFigFont{6}{7.2}{\familydefault}{\mddefault}{\updefault}{\color[rgb]{0,0,0}$w_2$}%
}}}}
\put(526,-1561){\makebox(0,0)[lb]{\smash{{\SetFigFont{6}{7.2}{\familydefault}{\mddefault}{\updefault}{\color[rgb]{0,0,0}$u_1$}%
}}}}
\put(6676,-2386){\makebox(0,0)[lb]{\smash{{\SetFigFont{6}{7.2}{\familydefault}{\mddefault}{\updefault}{\color[rgb]{0,0,0}$\Xi_\text{cav}(s)$}%
}}}}
\put(4126,-1636){\makebox(0,0)[lb]{\smash{{\SetFigFont{6}{7.2}{\familydefault}{\mddefault}{\updefault}{\color[rgb]{0,0,0}$v_1$}%
}}}}
\put(6001,-1636){\makebox(0,0)[lb]{\smash{{\SetFigFont{6}{7.2}{\familydefault}{\mddefault}{\updefault}{\color[rgb]{0,0,0}$w_1$}%
}}}}
\put(4801,-2986){\makebox(0,0)[lb]{\smash{{\SetFigFont{6}{7.2}{\familydefault}{\mddefault}{\updefault}{\color[rgb]{0,0,0}squeezer}%
}}}}
\put(2476,-2986){\makebox(0,0)[lb]{\smash{{\SetFigFont{6}{7.2}{\familydefault}{\mddefault}{\updefault}{\color[rgb]{0,0,0}beamsplitter}%
}}}}
\put(6526,-3361){\makebox(0,0)[lb]{\smash{{\SetFigFont{6}{7.2}{\familydefault}{\mddefault}{\updefault}{\color[rgb]{0,0,0}$G_{\text{dyn}}$}%
}}}}
\put(6826,-2986){\makebox(0,0)[lb]{\smash{{\SetFigFont{6}{7.2}{\familydefault}{\mddefault}{\updefault}{\color[rgb]{0,0,0}cavity}%
}}}}
\put(3226,-3361){\makebox(0,0)[lb]{\smash{{\SetFigFont{6}{7.2}{\familydefault}{\mddefault}{\updefault}{\color[rgb]{0,0,0}$G_{\text{static}}$}%
}}}}
\put(5101,-4186){\makebox(0,0)[lb]{\smash{{\SetFigFont{6}{7.2}{\familydefault}{\mddefault}{\updefault}{\color[rgb]{0,0,0}$\tau$}%
}}}}
{\color[rgb]{0,0,0}\put(6601,-2761){\framebox(975,900){}}
}%
\end{picture}%

\caption{Linear quantum feedback network of Figure \ref{fig:bs-dyn} redrawn in standard from as Figure \ref{fig:lqfn}.}
\label{fig:bs-dyn-equiv}
\end{center}
\end{figure}

We now describe the LQFN of Figure \ref{fig:bs-dyn}, which contains a cavity and  squeezer in a feedback loop resulting from interconnection with a beamsplitter. The total propagation delay around the loop is $\tau$, which we take to be small and send $\tau \to 0$. In order to determine an equivalent zero-delay limit model, following the general approach of section \ref{sec:gen-nw-zero}, we re-draw the network as shown in Figure \ref{fig:bs-dyn-equiv}.

As indicated in Figure \ref{fig:bs-dyn-equiv}, the in-loop system $G = G_\text{dyn} \triangleleft G_\text{static} $ is a dynamical Bogoliubov component obtained by cascading the beamsplitter, the (augmented) squeezer (which together form $G_\text{static}$), and the (augmented) cavity $G_\text{dyn}$.

The static part $G_\text{static}$ is described as follows. Because the beamsplitter has two inputs and two outputs, we augment the squeezer 
$S_\text{sq}$ (given by (\ref{squeezer})) by including a direct feed through channel ($v_1$ to $w_1$ in Figure \ref{fig:bs-dyn-equiv}). Because the squeezer is represented by a static Bogoliubov transformation expressed in doubled-up form, we express the beamsplitter in doubled-up form: $\tilde S_b = \Delta(S_b, 0)$. To be clear, the beamsplitter is described by
\begin{eqnarray}
v_1 &=& \alpha u_1 - \beta u_2
\nonumber \\
v_2 &=& \beta u_1 + \alpha u_2,
\label{bs-eqns}
\end{eqnarray}
where $\vert \alpha \vert^2 + \vert \beta \vert^2 =1$, $\alpha^\ast \beta = \beta^\ast \alpha$. Thus we have
\begin{equation}
S_b = \left[ \begin{array}{cc}
\alpha & -\beta
\\
\beta & \alpha
\end{array} \right],
\label{bs-1}
\end{equation}
and
\begin{eqnarray}
\tilde S_b &=& \Delta( S_b, 0)
\nonumber 
\\
&=&
\left[  \begin{array}{cccc}
\alpha & - \beta & 0 & 0 
\\
\beta & \alpha & 0 & 0
\\
0 & 0 & \alpha^\ast  & -\beta^\ast
\\
0 & 0 & \beta^\ast & \alpha^\ast
\end{array}
\right].
\label{bs-2}
\end{eqnarray}

The static component $G_\text{static}$  has  inputs $(u_1, u_2)^\top$ and outputs $(w_1, w_2)^\top$, and is given by the Bogoliubov matrix
\begin{eqnarray}
\tilde R  &=& \Delta \left(
\left[ \begin{array}{cc}
1 & 0
\\
0 & \cosh  r \end{array} \right] ,
\left[ \begin{array}{cc}
0 & 0
\\
0 & \sinh  r \end{array} \right] 
\right)
\tilde S_b
\nonumber
\\
&=&
\Delta(  R_-, R_+)
\label{G-static-1}
\end{eqnarray}
where
\begin{eqnarray}
R_- =  \left[ \begin{array}{cc}
\alpha  & -\beta
\nonumber  \\
\beta  \cosh  r  & \alpha \cosh r
\end{array} \right] , 
\\
R_+ =  \left[ \begin{array}{cc}
0 & 0 
\\
\beta^\ast  \sinh  r  & \alpha^\ast \sinh r
\end{array} \right].
\label{G-static-2}
\end{eqnarray}

The dynamic component $G_{\text{dyn}}$, with inputs $(w_1, w_2)^\top$ and outputs $(y_1, y_2)^\top$, is given by

\begin{eqnarray}
\left[ 
\begin{array}{c}
\dot{a} \\ 
\dot{a}^{\ast }
\end{array}
\right]  &=&\left[ 
\begin{array}{cc}
-(\frac{\gamma }{2}+i\omega ) & 0 \\ 
0 & -(\frac{\gamma }{2}-i\omega )
\end{array}
\right] \left[ 
\begin{array}{c}
a \\ 
a^{\ast }
\end{array}
\right] \notag \\
& &-\left[ 
\begin{array}{cccc}
0 & \sqrt{\gamma } & 0 & 0 \\ 
0 & 0 & 0 & \sqrt{\gamma }
\end{array}
\right] \left[ 
\begin{array}{c}
w_{1} \\ 
w_{2} \\ 
w_{1}^{\ast } \\ 
w_{2}^{\ast }
\end{array}
\right],   \notag \\
\left[ 
\begin{array}{c}
y_{1} \\ 
y_{2} \\ 
y_{1}^{\ast } \\ 
y_{2}^{\ast }
\end{array}
\right]  &=&\left[ 
\begin{array}{cc}
0 & 0 \\ 
\sqrt{\gamma } & 0 \\ 
0 & 0 \\ 
0 & \sqrt{\gamma }
\end{array}
\right] \left[ 
\begin{array}{c}
a \\ 
a^{\ast }
\end{array}
\right] +\left[ 
\begin{array}{c}
w_{1} \\ 
w_{2} \\ 
w_{1}^{\ast } \\ 
w_{2}^{\ast }
\end{array}
\right]  . \label{G-dyn-1}
\end{eqnarray}
Thus $A_- = -(\frac{\gamma}{2} +i \omega)$, $A_+=0$, $\tilde A = \Delta ( -(\frac{\gamma}{2} +i \omega), 0)$,
\begin{equation}
C_- = \left[ \begin{array}{c}
0 
\\
\sqrt{\gamma}
\end{array} \right], \ \ 
C_+ = \left[ \begin{array}{c}
0 
\\
0
\end{array} \right] , \ \ 
\tilde C = \Delta ( \left[ \begin{array}{c}
0 
\\
\sqrt{\gamma}
\end{array} \right], 0) .
\label{G-dyn-2}
\end{equation}
Also, $\Omega_-=\omega$, $\Omega_+=0$.

Now that we have a complete model for the in-loop system $G$, we may apply the formulas in section \ref{sec:gen-nw-zero} to obtain a zero-delay network model $N_0$. This involves first working out the Bogoliubov matrix in partitioned form:
\begin{eqnarray}
\hat S_{11} &=& \Delta( \alpha, 0), \ \ \hat S_{12} = \Delta( -\beta, 0), 
\nonumber \\
\hat S_{21} &=& \Delta( \beta \cosh r,  \beta^\ast \sinh r),
\nonumber \\
\ \ \hat S_{22} &=& \Delta( \alpha \cosh r,  \alpha^\ast \sinh r).
\label{hat-S}
\end{eqnarray}
We set $\alpha = \sqrt{\epsilon}$ and $\beta = \sqrt{1-\epsilon}$ to simplify some of the algebra. The network model is given as follows. We now use the equations (\ref{N-lft-S-I},\ref{N-lft-C-I},\ref{N-lft-A-I}) to determine the network parameters. The equivalent network Bogoliubov matrix is
\begin{equation}
\tilde S_0 = \Delta ( \sqrt{\epsilon} - \frac{1-\epsilon}{\mu} ( \cosh r - \sqrt{\epsilon} ), -\frac{(1-\epsilon) \sinh r}{\mu} ),
\label{S0-1}
\end{equation}
where
\begin{equation}
\mu = 1-2 \sqrt{\epsilon} \cosh r +\epsilon. 
\label{mu0-1}
\end{equation}
Next,
\begin{equation}
\tilde C_0 = -\frac{\sqrt{1-\epsilon} \sqrt{\gamma}}{\mu} \Delta( 1-\sqrt{\epsilon} \cosh r, \sqrt{\epsilon} \sinh r) ,
\label{C0-1}
\end{equation}
so that
\begin{eqnarray*}
C_0^- &=& -\frac{\sqrt{1-\epsilon} \sqrt{\gamma}}{\mu}(1-\sqrt{\epsilon}\cosh r), \\ 
C_0^+ &=& -\frac{\sqrt{1-\epsilon} \sqrt{\gamma}}{\mu} \sqrt{\epsilon} \sinh r .
\end{eqnarray*}
Now
$\tilde A_0=\Delta( A_0^-, A_0^+)$, where
\begin{equation}
A_0^-= -(\frac{\gamma}{2} +i\omega) - \frac{\sqrt{\epsilon} \gamma}{\mu} (\cosh r - \sqrt{\epsilon}), \ \ 
A_0^+ =  \frac{\sqrt{\epsilon} \gamma}{\mu} \sinh r
\label{A0}
\end{equation}
From this we compute
\begin{equation}
\Omega_0^- = \omega, \ \ \Omega_0^+ = i \frac{\sqrt{\epsilon} \gamma}{\mu} \sinh r .
\label{net-omega}
\end{equation}
We therefore see that not only does the network model $N_0 \in \mathcal{L}^{\text{Bog.}}(1)$ have a non-trivial static Bogoliubov term, it also has field couplings involving a creation operator $a^\ast$, and Hamiltonian terms involving $a^2$ and $(a^\ast)^2$.

Stability of the feedback system may be analyzed using the methods of section   \ref{sec:components-dyn-stab} or the  small gain theorem \cite{ZDG96}, \cite{DJ06}.

As a possible application, we note that the squeezing parameter of a DPA may be altered by placing it in-loop in a beam-splitter arrangement of this type \cite{GW09}.
%The loop gain is the product of the gains of the components going around the feedback loop, and stability is ensured if this number is less than one. In this example, this condition reads 
%$\sqrt{\epsilon}\, e^r < 1$.

\subsection{Dynamics from Feedback}
\label{sec:eg-2-dyn}

In this example we give an illustration from quantum optics showing that LQFNs involving only static components may give rise to dynamical behavior.  This dynamical behavior is due to a time delay in the feedback loop.
We consider the network shown in Figure \ref{fig:bs-dyn}, \cite{YK03a}, \cite{GJ08a}. This is a special case of the LQFN network of Figure \ref{fig:bs-dyn}, but with no squeezing and no cavity. The beamsplitter $S_b$ is given by (\ref{bs-eqns}) or \ref{bs-1}, with   $\alpha = \sqrt{\epsilon}$ and $\beta = \sqrt{1-\epsilon}$.

\begin{figure}[h]
\begin{center}

\setlength{\unitlength}{2368sp}%

\begingroup\makeatletter\ifx\SetFigFont\undefined%
\gdef\SetFigFont#1#2#3#4#5{%
 \reset@font\fontsize{#1}{#2pt}%
 \fontfamily{#3}\fontseries{#4}\fontshape{#5}%
 \selectfont}%
\fi\endgroup%
\begin{picture}(4844,3344)(579,-4583)
\put(1876,-4261){\makebox(0,0)[lb]{\smash{{\SetFigFont{7}{8.4}{\familydefault}{\mddefault}{\updefault}{\color[rgb]{0,0,0}$u_2$}%
}}}}
\put(3576,-4261){\makebox(0,0)[lb]{\smash{{\SetFigFont{7}{8.4}{\familydefault}{\mddefault}{\updefault}{\color[rgb]{0,0,0}$\tau$}%
}}}}
\thicklines
{\color[rgb]{0,0,0}\put(601,-2761){\vector( 1, 0){1725}}
}%
{\color[rgb]{0,0,0}\put(2326,-2761){\vector( 1, 0){1725}}
}%
{\color[rgb]{0,0,0}\put(2326,-2761){\line( 0, 1){1500}}
\put(2326,-1261){\line( 1, 0){3075}}
\put(5401,-1261){\line( 0,-1){3300}}
\put(5401,-4561){\line(-1, 0){3000}}
\put(2401,-4561){\line(-1, 0){ 75}}
\put(2326,-4561){\vector( 0, 1){1800}}
}%
\put(901,-2611){\makebox(0,0)[lb]{\smash{{\SetFigFont{7}{8.4}{\familydefault}{\mddefault}{\updefault}{\color[rgb]{0,0,0}$u_1$}%
}}}}
\put(3376,-2986){\makebox(0,0)[lb]{\smash{{\SetFigFont{7}{8.4}{\familydefault}{\mddefault}{\updefault}{\color[rgb]{0,0,0}$y_1$}%
}}}}
\put(2476,-1636){\makebox(0,0)[lb]{\smash{{\SetFigFont{7}{8.4}{\familydefault}{\mddefault}{\updefault}{\color[rgb]{0,0,0}$y_2$}%
}}}}
{\color[rgb]{0,0,0}\put(2926,-2161){\line(-1,-1){1200}}
}%
\end{picture}%

\caption{Beamsplitter feedback network with time delay $\tau$.}
\label{fig:bs-dyn-2}
\end{center}
\end{figure}
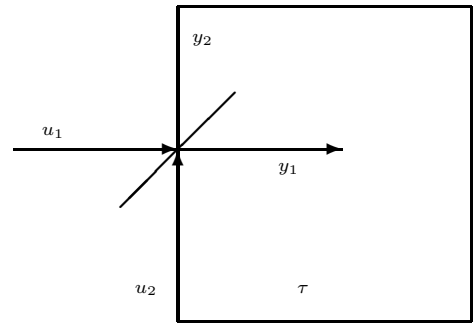

 Feedback in the network is defined by the constraint
$
u_2(t) = \Theta_\tau y_2(t) = y_2(t-\tau) ,
$
where $\tau > 0$ is the time taken for light to travel from the output to the input.

This network is a LQFN with $G^\epsilon=(\Delta(S_b,0), 0, 0) \in U(2)$.  With $\epsilon > 0$ fixed, 
the zero-delay network model $\tau \to 0$ is the system $N_0=\mathfrak{F}_l(G^\epsilon, I)=(\Delta(-I,0), 0, 0)\in U(1)$, with transfer function $N_0(s)=-1$, a trivial pass-through system with sign change (phase shift).  

%?? check for consistency with above 

Now if the reflectivity coefficient $\epsilon$  and the time delay are comparable, say $\tau= \epsilon/\gamma$, where $\gamma > 0$, then we obtain a dynamical model as $\epsilon \to 0$, \cite[Section 2.3]{BR04} (recall section \ref{sec:gen-nw-zero} above). Indeed, solving (\ref{bs-1}) and (\ref{bs-2}) in the frequency domain we find that the transfer function is
\begin{equation}
N^\epsilon(s)   =  \mathfrak{F}_l( G^\epsilon, \Theta_{\tau^\epsilon} )(s) = \sqrt{1-\epsilon}\, - \frac{  \epsilon e^{-s\epsilon/\gamma} }{ 1- \sqrt{1-\epsilon}\, e^{-s\epsilon/\gamma}  }
\label{bs-3}
\end{equation}
By L'Hopital's rule, we find that the limit transfer function is
\begin{equation}
N(s) = \lim_{\epsilon \to 0} N^\epsilon(s) =  1- \frac{\gamma}{ s+\gamma/2}  = \frac{s-\gamma/2}{s+\gamma/2}.
\label{bs-4}
\end{equation}
This transfer function corresponds to
a cavity $N = (I, \sqrt{\gamma}\, a, 0) =(I, \sqrt{\gamma}\, I, 0) \in \mathcal{L}^{\text{HP}}(1,1)$, where $a \in \mathcal{S}(1)$,  \cite{GZ00}, \cite{BR04}.
 Here, $\gamma$ plays the role of the coupling strength between the trapped cavity mode and the external free field.

This example shows that $U(2)$ is not closed under this type of physically natural approximation process (since the limit belongs to $\mathcal{L}^{\text{HP}}(1)$ which is outside $U(2)$), while $\mathcal{L}^{\text{Bog.}}(2)$ is closed (since it contains both $U(2)$  and 
$\mathcal{L}^{\text{HP}}(1)$).

\section{Conclusion}
\label{sec:conclusion}

We have shown how to extend linear quantum dynamical network theory to include static Bogoliubov components (such as squeezers).
This unified framework accommodates squeezing components which are important in quantum information applications. We provided tools for describing network connections and feedback using generalizations of linear fractional transformations and the series product,  \cite{GJ08}, \cite{GGY08}, \cite{YK03a},  \cite[Chapter 10]{ZDG96}. We have also defined input output maps and transfer functions within this linear quantum network theory, and shown how they can be used in applications. Finally, we explained the natural group structure arising from the series product.
% The general linear quantum feedback network theory provides a framework for use in quantum technology that is analogous to well established classical linear circuits and systems theory and its application in classical feedback control and electrical engineering.

\begin{acknowledgments}
We wish to acknowledge the support of the Australian Research Council. Thanks also to Guofeng Zhang, Robin Hudson, Rolf Gohm and Sebastian Wildfeuer for their comments.
\end{acknowledgments}

\appendix

%%%%%%%%%%%%%%%%%%%%%%%%%%%%%%%%%%%%

\bibliographystyle{plain}

%\bibliography{mjbib2004}

\begin{thebibliography}{10}
\bibitem{footnote}
To see this, let $E\geq 0$ be a complex $k\times k$ matrix with $E_{11}=0$. Set $u\left(
t\right) =\left( tx_{1},x_{2},\cdots ,x_{k}\right) ^{\top }$ then we have $%
0\leq u\left( t\right) ^{\dag }Eu\left( t\right) =2t$Re$%
\sum_{j>1}x_{1}E_{1j}x_{j}+\sum_{j,k>1}x_{j}^{\ast }E_{jk}x_{k}$. However,
the only way to guarantee this inequality for all real $t$ is to require that
Re$\sum_{j>1}x_{1}E_{1j}x_{j}=0$. Replacing $t$ with $it$ shows that Im$%
\sum_{j>1}x_{1}E_{1j}x_{j}$ likewise vanishes. As this must be true for
arbitrary $x_{j}$, we conclude that $E_{1j}=E_{j1}^{\ast }=0$ for all $j>1$.
\end{thebibliography}

\end{document}